\documentclass[10pt]{iopart}

\usepackage{iopams}  
\usepackage{cite}  
\usepackage{graphicx}
\usepackage[breaklinks=true,colorlinks=true,linkcolor=blue,urlcolor=blue,citecolor=blue]{hyperref}
\usepackage{braket}
\usepackage{units}

\usepackage[utf8]{inputenc}
\usepackage[T1]{fontenc}
\usepackage{times}
\usepackage[american]{babel}

\newcommand{\db}[2][]{\text{d}^{#1}#2}
\newcommand{\text}[1]{\textnormal{#1}}
\renewcommand{\vec}[1]{\bi{#1}}

\newcommand{\avr}[1]{\braket{#1}}

\newcommand{\abs}[1]{|#1|}
\newcommand{\al}{\ensuremath{\alpha}}
\newcommand{\be}{\ensuremath{\beta}}

\begin{document}

\title[Microscopic modeling of gas-surface scattering. I. Combined MD-RE approach]{Microscopic modeling of gas-surface scattering. I. A combined molecular dynamics-rate equation approach}


\author{A Filinov$^{1,2,3}$, M Bonitz$^1$ and  D Loffhagen$^2$}
\address{$^1$ Institut f\"ur Theoretische Physik und Astrophysik,
Christian-Albrechts-Universit\"at, Leibnizstr. 15, D-24098 Kiel, Germany}
\address{$^2$ INP Greifswald e.V., Felix-Hausdorff-Str.~2, D-17489 Greifswald, Germany}
\address{$^3$ Joint Institute for High Temperatures RAS, Izhorskaya Str.~13, 125412 Moscow, Russia}
\ead{filinov@theo-physik.uni-kiel.de}


\begin{abstract}

A combination of first principle molecular dynamics (MD) simulations with a rate equation 
 model (MD-RE approach) is presented 
to study the trapping and the scattering of rare gas atoms from metal surfaces. 
The temporal evolution of the atom fractions that are either adsorbed or scattered into the continuum is investigated in detail. 
We demonstrate that for this description one has to consider trapped, quasi-trapped and scattering states, and  
present an energetic definition of these states. 
The rate equations contain the transition probabilities between the states. We demonstrate how these rate equations can be derived from kinetic theory. Moreover, we present a rigorous way to determine the transition probabilities from a microscopic analysis of the particle trajectories generated by MD simulations. 
Once the system reaches quasi-equilibrium, the rates converge to stationary values, and the subsequent thermal adsorption/desorption dynamics is 
completely described by the rate equations without the need to   perform further time-consuming  MD simulations. 

As a proof of concept of our approach, MD simulations for argon atoms interacting with a platinum (111) surface are presented. A detailed deterministic trajectory analysis is performed, and the transition rates are constructed. The dependence of the rates on the incidence conditions and the lattice temperature is analyzed. 
Based on this example, we analyze the 
time scale of the gas-surface system to approach the quasi-stationary state. 
The MD-rate equation model  has great relevance for the plasma-surface modeling as it makes  
an extension of accurate simulations to long, experimentally relevant time scales possible. 
Its application to the computation of atomic sticking probabilities is given in the second part (paper II).
\end{abstract}

\noindent{\it Keywords}: plasma-surface modeling, low-temperature plasma, gas-surface interaction, adsorption and scattering of neutral particles, molecular dynamics, rate equations, transition rates, quasi-equilibrium, residence time, argon atom, platinum surface

\submitto{\PSST}                            

\maketitle

\ioptwocol

\section{Introduction}\label{intro}
Low-temperature plasma physics has seen remarkable progress over the last decade. This concerns both fundamental science studies and technological applications ranging from etching  of solid surfaces to plasma chemistry and plasma medicine. In each of these applications, the contact between particles of the plasma and a  solid plays a crucial role. This contact is very complex and includes a large variety of fundamental physical processes such as secondary electron emission, sputtering, neutralization and stopping of ions and adsorption and scattering of neutral particles as well as chemical reactions. In the majority of previous studies in low-temperature plasma physics, these processes have been omitted or treated phenomenologically. For example, in many state of the art kinetic simulations based on the Boltzmann equation, e.g. \cite{hagelaar_2005,donko_2016} or particle in cell (PIC) simulations, e.g. \cite{ebert_2016,becker_2017} neutrals are treated as a homogeneous background, and their interaction with surfaces is not included in the description. The fact that energetic neutrals maybe crucial for secondary electron emission was demonstrated in PIC simulations of Derszi \textit{et al.} where neutrals above a threshold energy of 23eV were traced \cite{derszi_2015}. They also took into account the effect of the plasma-induced surface modification by using modified cross sections \cite{phelps_1999}. Similarly, Li \textit{et al.} studied the effect of surface roughness on the field emission by including a phenomenological geometric enhancement factor \cite{li_2013}.
Not surprisingly, a better understanding of plasma-surface interaction, which would lead to predictive capability, has been recognized to be a major bottleneck for further progress in the field~\cite{plasma-road-map-17, doe-report-17}.

The goal of the present work is to present a microscopic theory-based approach to a specific problem of plasma-surface interaction: the scattering, adsorption and sticking of rare gas atoms from a plasma at a metal surface. This is expected to be particularly important at low pressures where the neutral fraction of heavy particles is by far dominating over the ionized component. Our approach presents a combination of microscopic modeling within a Langevin molecular dynamics (MD) approach with an analytical model, formulated as rate equations for relevant surface states. This coupled approach has the advantage of opening the way towards \textit{ab initio} based long-time simulations, as we will explain in detail below. 
 
We start with a brief overview on previous theoretical works treating the interaction of gas atoms with solid surfaces and summarize their strengths and limitations.
The energy transfer and the scattering of rarefied gases from different surfaces have been the subject of multiple studies.
Much effort has been devoted to determine an accurate scattering model for gas atoms at a solid surface and a realistic description of the involved collision processes. 
Already Maxwell~\cite{Max} proposed  a simple model in his studies of gas–surface interactions, where the scattered gas atoms are separated into two fractions: one that is reflected specularly and exchanges no energy, and a second one that is accommodated  completely and, eventually, desorbs with an equilibrium distribution. Other studies have been focused on 
the evaluation of the thermal accommodation coefficients~\cite{robert}. 

The angular distributions for different gas atoms (helium, neon, argon, krypton, xenon, and deuterium) scattered by a single-crystal tungsten (110) surface has been intensively analyzed by Weinberg and Merrill~\cite{Weinberg}. A direct measurement of the velocity distribution of argon atoms scattered from poly-crystalline surfaces has been in the focus of the work of Janda \etal~\cite{Ar}. These measurements allowed to reveal a dependence of the average kinetic energy of scattered argon atoms on the average incident kinetic energy and surface temperature. 

A theoretical treatment of the interaction of Ar atoms with a self-assembled monolayer on Ag(111) in terms of the stochastic scattering theory, including direct scattering, trapping, and desorption, has been reported by Fan and Manson~\cite{Mans}. Gibson \etal~\cite{gibson} presented a detailed study of Ar scattered from an ordered 1-decanethiol–Au(111) monolayer. 

All these detailed experimental and theoretical studies have proven that 
the use of 
atomic probes as scattering projectiles can be a useful experimental tool for studies of structure and dynamical properties of surfaces. The scattered intensities can be measured as a function of the final translation energy, scattered and incident angles. Furthermore, such analyses provide important information on surface corrugation and temperature effects. 

Presently, there exist three main theoretical research directions in this area. The first one is based on kinetic theory where important contributions are due to 
Kreuzer and Teshima 
\cite{kreuzer_prb81} and Brenig \cite{brenig_zphysb82, brenig_pscr87}.  More recently Bronold \textit{et al.} studied the sticking of electrons at a dielectric surface \cite{bronold_prl15} and the neutralization of ions at a gold surface \cite{pamperin_prb15}. Overall, the resulting kinetic equations provide a powerful semi-analytical tool to compute (mainly) stationary properties of gas atoms or charged particles scattering from a surface within suitable many-body approximations.

The second direction is based on \textit{ab initio} quantum simulations, most importantly Born-Oppenheimer density functional theory (DFT) or time-dependent DFT (TDDFT). The main advantage of the latter is that the dynamics of the electrons and the internal excitation of adsorbed atoms and molecules can be incorporated on a full quantum level. Among recent applications of TDDFT, 
we mention the analysis of the chemical reaction dynamics of hydrogen on a silicon surface \cite{brenig-pehlke_08} and the energy loss (stopping power) of ions on graphene and boron nitride sheets \cite{zhao_prb_15}. An alternative \textit{ab initio} approach is based on nonequilibrium Green functions (NEGF) and is advantageous when electronic correlations in the surface are of importance, e.g. \cite{bonitz_qkt, schluenzen_cpp_16}. Recently NEGF simulations of the stopping power of hydrogen and helium ions on correlated two-dimensional materials were presented \cite{balzer_prb_16}. 

The third approach is semiclassical MD simulations that are extensively used in surface science. Here, the quality depends on the accuracy of effective pair potentials (or force fields) whereas details of the electron dynamics are not resolved. As examples of recent applications, we mention the simulation of metal cluster growth and diffusion \cite{bonitz_cpp_12,abraham_jap_16} and of the dynamics of bi-metallic clusters~\cite{abraham_cpp_18}.

The temporal evolution of the atoms trapped near the surface is of particular interest for the understanding of the adsorption and scattering of neutral atoms. 
Their equilibration kinetics is 
of fundamental importance for the 
understanding of the dependence of the sticking probabilities on 
the energy, incident angle and lattice temperature.
However, these time dependences are typically out of the range 
of both kinetic theories and \textit{ab initio} quantum simulations. 
While the former usually concentrate on stationary properties,  TDDFT and NEGF simulations are computationally extremely expensive and are limited to very short times of the order of a picosecond and small spatial scales.
Therefore, MD simulations represent the only approach capable of resolving the dynamics on sufficiently long time and spatial scales being of experimental relevance. 
Although the typically required time step in these simulations is well below one femtosecond, 
recent progress in computer power made it possible to 
 investigate the relevant surface effects on a microscopic level reaching simulation times of several hundreds of picoseconds.

In the present work (paper I and paper II) we perform extensive MD simulations to study the scattering and sticking of argon atoms at a platinum (111) surface in a time-resolved fashion. 
In order to analyze the sticking and thermalization behavior, we introduce a novel approach: the trajectories of gas atoms that are near the surface are sub-divided into three classes: trapped (T), quasi-trapped (Q) and continuum (C) states. We demonstrate that these 
states 
are the relevant observables to analyze the sticking problem with excellent statistics, high accuracy and temporal resolution. 
Furthermore, we demonstrate that the fractions of atoms in trapped, quasi-trapped and continuum states obey a simple system of coupled rate equations. 
Its solution allows us to significantly reduce the computational cost that is otherwise spent on the temporal resolution of individual particle trajectories. 
Moreover, we demonstrate how the transition rates between the three states can be accurately extracted from our MD simulations transforming the rate equations into an, in principle, exact description. The accuracy of the latter is only limited by the accuracy of the used pair potentials. Finally, our analysis of the temporal evolution reveals that after a characteristic equilibration time these transition rates become stationary. This means that 
the rate equations become sufficient to study the dynamics of the systems 
for longer times 
without further need of MD simulations. This provides the potential to significantly extend the temporal and spatial scales of the simulations without compromising the accuracy.

The main goal of the present paper is to introduce this new combined molecular dynamics-rate equation (MD-RE) approach in detail 
and to test it thoroughly on a representative example: the scattering of argon atoms at a platinum (111) surface. Even though for low-temperature plasma applications other metals are more common, we chose platinum as a test case. 
Here extensive experimental and theoretical data are available for comparison, which allow us to critically assess the validity and limitations of our model. 
A detailed analysis of the argon sticking probability, its  dependence on temperature, incident energy and angle is 
presented in a separate paper (``paper II'', cf.~Ref.~\cite{paper2}).

The paper is organized as follows. In section~\ref{model1} we introduce the microscopic model of the gas-surface interaction. 
The setup of the MD simulations is explained in section~\ref{model}. 
The tracking of the particle trajectories provides a direct access to time dependencies of the energy loss distribution functions 
and the average momentum (kinetic energy). The time-resolved evolution of the trapped, quasi-trapped, and continuum states and its dependence on 
the lattice temperature and incident angle is discussed in detail in section~\ref{model-states}. 
In sections~\ref{s:derivation} and \ref{rateSec} we introduce the rate equation model, 
which allows us to reproduce the MD results of section~\ref{model-states}, and, moreover, has a potential to extrapolate these data to much longer times being not accessible by
usual MD simulations  [see subsection~\ref{ss:analytics}]. The conclusions are given in section~\ref{s:conclusion1}.

\section{Effective gas-surface interaction}\label{model1}

In a common approach, the global interaction potential 
$V_i^{g}$ 
of the $i$-th gas-atom (a) with the surface (s) is decomposed into a sum of pair potentials $V^{as}$ according to 
\numparts
\begin{eqnarray}
V_i^{g}(Z) &=&\sum_{j=1}^{N_{s}} V^{as}(r_{ij}),\label{Uglob}\\
V^{as}(r) &=&\nu_0 e^{-\al r} -\frac{C_6} {r^6}  -\frac{C_8}{r^8}\,,
 \label{Uglob1}
\end{eqnarray}
\endnumparts
where $r_{ij}=\abs{\vec r^a_i -\vec r^s_j}$ is the distance between the $i$-th gas-atom and the $j$-th atom of the surface, 
 $N_{s}$ denotes the number of surface atoms,  and the set of coefficients $\{\nu_0,\al,C_6,C_8\}$ used for the parametrization of the Ar-Pt pair potential is specified in Table~\ref{tab1}.
The distance to the surface, $Z = z_i$, in Eq.~(\ref{Uglob}) is treated as a parameter 
\begin{eqnarray}
 r_{ij}(Z)=\sqrt{x_{ij}^2+y_{ij}^2+(Z-z_j)^2} \,,
\end{eqnarray}
while the lateral atom position $(x_i, y_i)$ is kept fixed to analyze the dependence of the gas-surface interaction on the adsorption site (see below).  
However, this lateral position 
can be varied for the  calculation of the potential energy surface (PES).

Such type of global interaction potential and its decomposition into  pairwise terms for the  Ar-Pt(111) system 
has been recently reconstructed using the periodic DFT
approach~\cite{Leonard}. Similar analyses have been performed for the interaction between an argon atom and gold surfaces~\cite{Greiner}. In general, the computation of such interaction potentials is still a challenging task, but 
it 
follows a standard scheme. \textit{Ab initio} calculations are performed for relatively small surface clusters, where the reconstructed pair potential~(\ref{Uglob1}) can be cross-checked to compare their performance  with state-of-the-art 
van der Waals-corrected  
periodic DFT approaches. As a result, accurate pair potentials suitable for molecular dynamics simulations are derived. 
The parametrization values used 
for the present Ar-Pt(111) system 
are listed in table~\ref{tab1}.

\begin{table} 
\caption{Parameters of the Ar-Pt(111) interaction potential used in Eq.~(\ref{Uglob1}), taken from Ref.~\cite{Leonard}: $\nu_0=3485.40$ eV, $\alpha=3.30$ \AA$^{-1}$, $C_6=64.92$ eV$\cdot$\AA$^6$, $C_8=0.0$ eV$\cdot$ \AA$^8$.
The equilibrium distance to the plane $z=0$ and the interaction energy $E_0=V^g(r_0)$ at the different adsorption sites [see Fig.~\ref{fig:UArPt}] are evaluated for a ($7\times 4\times 2$) slab [number of unit cells in the X, Y and Z directions, each with 6 Pt atoms] at 
the surface temperature $T_s=0$. The values $r_0^\star$ and $E_0^\star$ correspond to the relaxed lattice when the positions of a few upper layers are optimized (see text).}
  \label{tab1}
 \begin{tabular}{c c c c c}
 \hline
 \hline
 & hcp & fcc & atop & bridge\\
 \hline
$r_0$ [\AA]  & 3.3576 & 3.3576  & 3.5058  & 3.3788\\
$E_0$ [meV]  & -76.8951 & -76.8876  & -68.8091  & -75.7618\\
\hline
$r_0^\star$ [\AA]  & 3.2412 & 3.2412  & 3.3894  & 3.2624\\
$E_0^\star$ [meV]  & -78.1302 & -78.1246  & -69.8872  & -76.9785\\
  \hline
 \hline
 \end{tabular}
 \end{table}

\begin{figure}
  \begin{center} 
\includegraphics[width=0.49\textwidth]{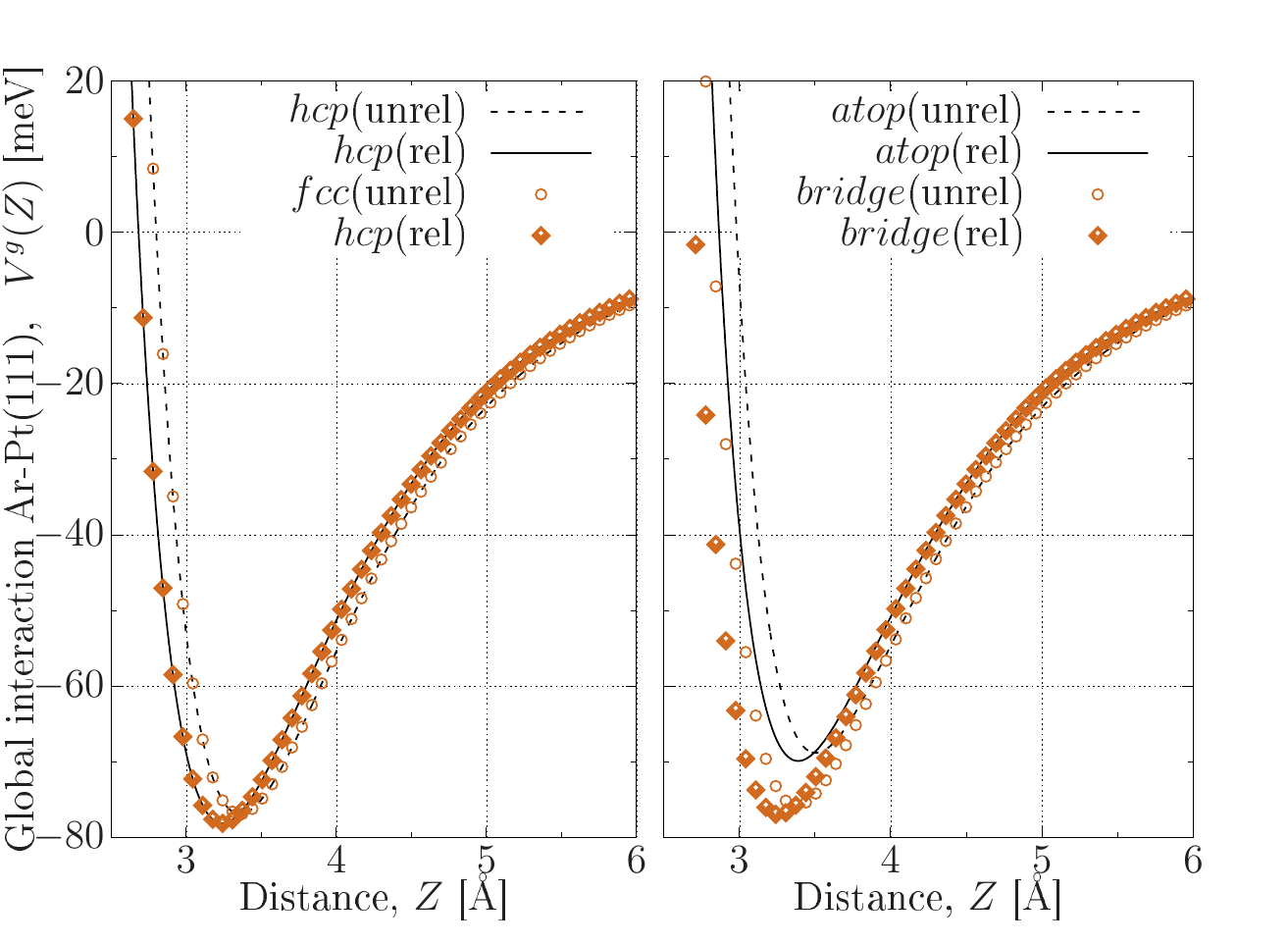}\\
\includegraphics[width=0.49\textwidth]{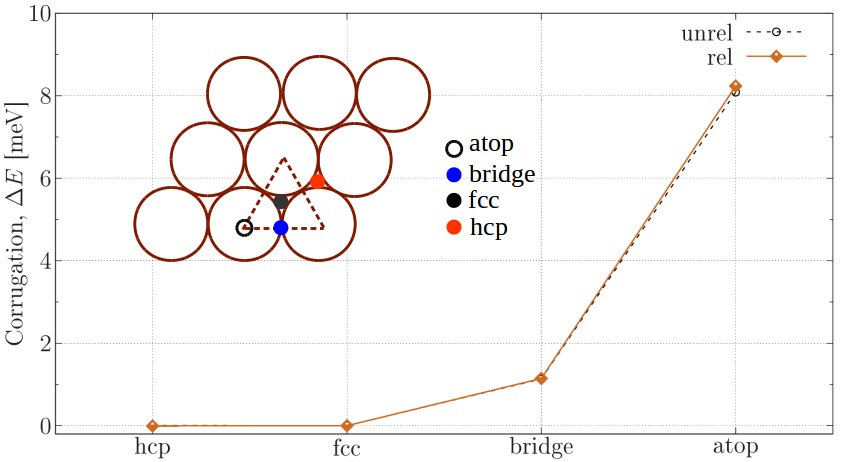}
  \end{center}
  \vspace{-.60cm}
  \caption{Top: Global interaction potential $V^{g}(Z)$ for Ar-Pt(111) for a set of adsorption sites specified by the lattice symmetry. 
  Unrelaxed (``unrel'') and relaxed (``rel'') lattices are compared (see text).
  Bottom: Corrugation of the global potential $\Delta E$ plotted with respect to the energy of the hcp-site.}
  \label{fig:UArPt}
\end{figure}

The obtained global Ar-Pt surface interaction potential is presented in figure~\ref{fig:UArPt} as a function of the height $z$ above the surface plane $z=0$. For the unrelaxed lattice the uppermost layer is placed at $z=0$.  The lateral position of the Ar atom has been varied in the $x$-$y$-plane to analyze the corrugation of the potential energy surface. Several sites according to the lattice symmetry (``atop'', ``bridge'', ``fcc'' and ``hcp'') have been chosen, as also specified in figure~\ref{fig:UArPt} for the fcc(111) lattice. The corresponding potential energy minima and distance to the surface are given in table~\ref{tab1}.
Their difference with respect to the energy of the fcc site is 
additionally plotted in figure~\ref{fig:UArPt}. It characterizes the corrugation of the PES. 

In figure~\ref{fig:UArPt} (top) we compare the results obtained for the ``ideal'' (unrelaxed) lattice with the optimized (``relaxed'') lattice. The latter was reconstructed at 
the surface temperature $T_s=0$ by minimization  of the total energy of the ($7\times 4\times 2$) Pt slab.   
It accounts for a shift of a few upper layers to more negative values of the $z$-coordinate, where the bottom three layers are kept fixed at the same position as in the unrelaxed lattice. 
This fact can be seen in figure~\ref{fig:UArPt} by a shift of the potential energy minima  (see the curves labeled ``rel'') to smaller heights relative to the plane $z=0$. The estimated binding energy of $-78$\,meV is in good agreement with the total adsorption energy of about $-80$\,meV reported in the experimental work of Head-Gordon \textit{et al.}~\cite{gordon}. This value is also used as the reference for the optimization of the empirical potentials aimed to reproduce the experimental results  on the vertical Pt(111)-Ar harmonic vibrational frequency of 
$\omega(r_0)\sim 5$\,meV and on the trapping, desorption and scattering data~\cite{Lah,Kul,Svan}. For an overview of the most commonly used empirical potentials we refer to the recent work of L\'{e}onard \textit{et al.}~\cite{Leonard}.

In order to perform a diffusive motion on the frozen surface, i.e., lattice atoms are fixed in their equilibrium position, 
the kinetic energy of the adsorbate atom should exceed the energy barrier $\Delta E$ at the bridge site. However, no such restriction applies at finite temperatures as the adsorbate atoms can continuously exchange energy with the lattice atoms. In the following, three different lattice temperatures are simulated: 
$T_s=80$, 190, and $300$\,K, corresponding to 6.90, 16.38, and 25.7\,meV, respectively. 
Hence,  the thermal energy supplied from the lattice is significant to overcome the energy barrier for the inter-site hops in all cases. 

\section{MD simulation of the lattice at finite temperature}\label{model}

We perform deterministic molecular dynamics simulations to study the Ar-Pt(111) system. 
A $(7 \times 4 \times 2)$ platinum crystal slab of $336$ atoms (consisting of 6 atomic layers) is used. The sample is divided into three parts. 
Three lower layers form a static crystal. The atoms in this part are frozen at their equilibrium positions and serve as a basement for the dynamical upper layers. The three uppermost layers are an active zone of the crystal. They interact dynamically with the incoming gas atoms. Two bottom layers in the active zone consist of the atoms which are used as a boundary thermostat to realistically treat the removal of energy from the active zone. In this way the excess kinetic energy can dissipate from the active zone. This removal of excess kinetic energy is simulated by restoring the temperature of the heat bath atoms using Langevin dynamics~\cite{mdref}. The Langevin term  keeps the kinetic energy of the lattice atoms at the value specified by  the lattice temperature $T_s$. Finally, the Langevin term was switched off for the uppermost layer, 
and only dynamical correlations due to the gas-lattice (Ar-Pt) and lattice atoms (Pt-Pt) binary interactions  are retained.

The system of $N=N_{a}+N_{s}$ atoms is described by the potential energy
\begin{eqnarray}
V=\sum\limits_{i=1}^{N_{a}}\sum\limits_{j=1}^{N_{s}} V^{as}(|\vec r_{i}-\vec r_{j}|) +\sum\limits_{i<j}^{N_{s}} V^{ss}(|\vec r_{i}-\vec r_{j}|)
\end{eqnarray}
where $N_a$ is the number of gas atoms. The interactions between the gas atoms are neglected, 
i.e., $V^{aa}(r)=0$, assuming sufficiently low gas density. This situation is quite typical for low-temperature and low-pressure plasmas. Also, in our analysis we concentrate on the influence of the incidence conditions and lattice temperature on the adsorption process.  Effects due to pre-existing adsorbed atoms, i.e. related to a finite coverage of the surface are a topic on its own and will be studied elsewhere. This will also require an extension of the rate-equation model presented in Sec.~\ref{s:derivation} to an inhomogeneous system.

Effective atom-atom pair potentials are used 
for both Ar-Pt [see Eq.~(\ref{Uglob1})] and Pt-Pt interactions.
The interactions between the Pt atoms are modeled using the modified embedded atom potential, which quite accurately reproduces the spectrum of the transverse and longitudinal phonons~\cite{Ptphonon}. 
In the directions parallel to the surface ($x$-$y$ plane) periodic boundary conditions are applied.

At the beginning of the simulation, i.e., before introducing the Ar atom, 
the crystal is allowed to equilibrate to the surface temperature $T_s$. This procedure takes about $6$\,ps. 
The equilibration is monitored by the instantaneous lattice kinetic energy. After the equilibration phase ($t > 6$\,ps) the distribution functions of the 
tangential ($E^{\parallel}_k$)
and normal ($E^{\perp}_k$)
components of the kinetic energy 
for the atoms in the three dynamical layers have been evaluated. They quite accurately reproduce the expected 
(one-dimensional and two-dimensional) 
Boltzmann distribution at the temperature $T_s$
\begin{eqnarray}
P(E_k^{\perp}) &=& \frac{1}{\sqrt{\pi E_k^{\perp} T_s}} \exp{(-E_k^{\perp}/T_s)},\label{BEz}\\
P(E_k^{\parallel}) &=& \frac{1}{T_s} \exp{(-E_k^{\parallel}/T_s)}\label{BExy}.
\end{eqnarray}
This procedure confirms the correct implementation of the heat bath via Langevin molecular dynamics.

After the lattice has approached steady state, an Ar atom is introduced at a height $z=20$\,\AA \ above the surface outside the 
cut-off radius of the potential $V^{as}$. 
The trajectories are obtained by integrating the classical equations of motion using 
a fourth-order propagator algorithm~\cite{mdref}, to accurately account for the effect of the random force (the Langevin term). The integration time step is fixed at $0.06$\,fs. Trapping probabilities, energy exchange calculations and energy distribution functions are evaluated based on samples of $1000-5000$ trajectories. During an ``elementary'' event, the impinging gas atom interacts with all atoms within the cut-off radius $r^c_{\text{Ar-Pt}}\sim 10$\,\AA. A simulated trajectory is stopped when 
(i) the gas atom leaves the surface after undergoing one or more collisions with the surface 
(called ``bounces'') and attains a distance above the surface greater than $z^c=r^c_{\text{Ar-Pt}}$  
and (ii) the gas atom experiences more than $n_b = 40$ bounces.

The value $n_b=40$ 
of the number of bounces 
was chosen empirically. 
By tracking the temporal evolution of trajectories with $n_b \leq 40$, 
we observe 
in most cases 
 the thermalization of the adsorbate atoms to the lattice temperature and convergence of the energy distribution functions to their quasi-stationary form 
 on the corresponding time scale. 
 In particular, we analyzed the accommodation of the parallel and perpendicular momentum components. 
 Our MD simulations at low and high temperatures reproduce the general trend of a slower accommodation of the parallel momentum component, as 
 it was first pointed out by Hurst \textit{et al}~\cite{Auerbach}
 and confirmed in many other analyses~\cite{collins,smith1,smith2,Leonard}. This thermalization analysis for the system Ar on Pt (111) 
 is presented in detail in paper~II\cite{paper2}.

\section{Time-resolved trapped, quasi-trapped and scattered fractions}\label{model-states}

In many physical systems a strong chemical bonding to the surface is the dominant trapping mechanism. In contrast, in the present system 
the trapping process occurs due to the phys\-isorption potential well described by a van der Waals-type potential, cf.~Eq.~(\ref{Uglob1}). Depending on the lattice temperature, the trapped particles desorb after a finite residence time [see subsection~\ref{ss:analytics}]. The central question is whether this time is sufficient for the adsorbate atoms to equilibrate with the surface so that they eventually desorb with a  (quasi)equilibrium energy distribution function. We underline that this is not an  academic question but one of practical importance. 
Indeed,  the theoretical description is expected to simplify significantly in cases where thermalization is observed.

This problem was first put forward by Maxwell in his studies of gas-surface interactions~\cite{Max} and was further taken up by Knudsen~\cite{Knud}. The basic concept is based on the thermal accommodation and the efficiency of energy exchange at the gas-surface interface. The latter crucially depends on both the incident gas parameters and surface characteristics. In the special case of complete accommodation the desorbed particles leave the surface with a distribution function specified by the Knudsen flux~\cite{Mans}. This type of assumption is frequently used to explain  experimental data for the sticking probabilities of the adsorbate and for the distribution functions of energy, momentum and flux vector of the thermally desorbed atoms.

One of our goals is to test this assumption by microscopic modeling of the gas-surface interactions and analyze the equilibration kinetics. Using realistic calculations based on the microscopic model introduced in sections~\ref{model1} and~\ref{model},  we aim at reproducing the available experimental data for the sticking probabilities~\cite{expPt} and at providing a general framework for the analysis of the temporal evolution of the atom states localized near the surface. While we treat the scattering classically, quantum-mechanical effects are included 
by means effective interaction potentials, which are constructed from ground-state DFT calculations~\cite{Leonard}. 
Inelastic effects observed in the scattering events at large energies and high surface temperature are fully taken into account by the Langevin MD scheme~\cite{mdref}.

\subsection{Classification of particle trajectories: trapped, quasi-trapped, and scattering states}\label{ss:classification}

\begin{figure}
  \begin{center} 
  \hspace{-0.cm}\includegraphics[width=0.5\textwidth]{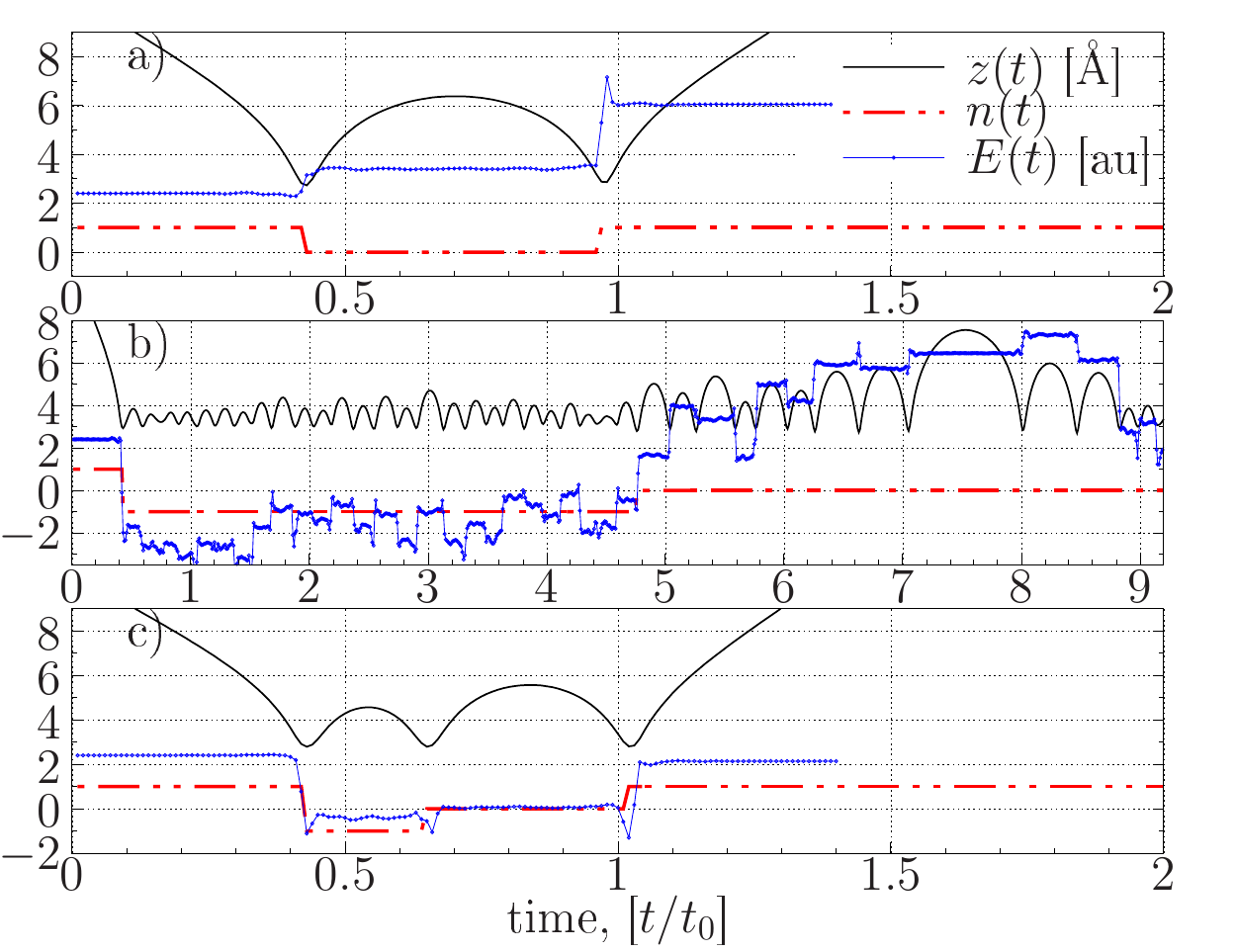} 
  \end{center}
  \vspace{-.80cm}
  \caption{Temporal evolution of the surface states $n(t)$ (see main text) along three dynamical trajectories based on the sign of the total energy $E(t)$ and the trapping condition, $E_{\text{k}}^{\perp} + V <0$. The curve $z(t)$ shows the distance of the  particle from the surface. The lattice temperature equals $T_s=300$\,K, and the incidence parameters of the atoms are $E_i=36.34$\,meV and $\theta=60^\circ$.}
  \label{fig:nstates}
  \end{figure}

The scattering from the surface is modeled by using mono-energetic gas atoms with fixed values of incident kinetic energy $E_{i}$
and angle $\theta$. A single collision with the surface introduces a transition from an initial momentum, $\vec p_i$, to a new momentum state $\vec p_f$. The final states ``f'' are distinguished  by the surface normal, $p_{f}^{\perp}$, and parallel, $p_{f}^{\parallel}$,  components, i.e. $\textbf{p}_f=\textbf{p}_{f}^{\perp}+\textbf{p}_{f}^{\parallel}$. Correspondingly, the kinetic energy 
$E_{\text{k}}$ of a particle with momentum \textbf{k} is split into two orthogonal contributions, 
\numparts
\begin{eqnarray}
 E_{\text{k}} &=& E_{\text{k}}^{\perp} + E_{\text{k}}^{\parallel}\,,\\
 E_{\text{k}}^{\perp} &=& \frac{(p^{\perp})^2}{2m}\,,\\
 E_{\text{k}}^{\parallel} &=& \frac{(p^{\parallel})^2}{2m}
 \,.
 \end{eqnarray}
 \endnumparts
In addition, every particle moves in the potential landscape of the surface atoms that is characterized by the local surface binding (physisorption) potential, $V$, 
giving rise to the total energy 
\begin{eqnarray}
 E(\vec r)=E_{\text{k}}(\vec r)+V(\vec r)\,.
 \label{eq:etot}
\end{eqnarray}
In the following, the dependence of $\{E,E_{\text{k}},V\}$ on the local atom position $\vec r$ is not 
explicitly specified. 
In addition, 
all energies and their corresponding distribution functions are evaluated at a minimum distance $\vec r^\star$ from the surface, where the total energy of a gas atom is approximately conserved. 
The temporal evolution of the total energy $E(t)$ is shown in figure~\ref{fig:nstates}.  Here and 
in the following, 
we use the value $t_0=10^3 \cdot a_0 \sqrt{m_{\mathrm{Ar}}/{E_\mathrm{h}}}=\unit[6.53]{ps}$ as a time unit, 
where $a_0$ is the Bohr radius, $m_{\mathrm{Ar}}$ denotes the atomic mass of the argon atoms, 
and $E_\mathrm{h} =\unit[27.211]{eV}$ is the Hartree energy. 
It becomes clear that plateau regions of 
$E(t)$ 
are well separated from the energy "jumps"  corresponding to the inelastic collision processes. 
During this inelastic process, there is a strong energy exchange with the surface atoms and, therefore, the evaluation of Eq.~(\ref{eq:etot}) becomes meaningless.

After interaction with the surface a fraction of particles is scattered back, whereas another fraction is (temporally) trapped. 
A classification 
of the states 
is straightforward by analyzing the particle energy.  
\begin{description}
  \item[I. Scattering states:] the condition to leave the surface due to the momentum exchange is 
    $E_{\text{k}}^{\perp} + V >0$. 
  The back-scattered (unbound) particles are referred to as ``continuum'' (\textbf{``C''}) states.
  \item[II. Bound states:] particles with 
    $E_{\text{k}}^{\perp} + V <0$ 
  remain localized near the surface.  The localization depends solely on the normal component, while the parallel component can be arbitrary. Therefore, depending on the sign of the total energy~(\ref{eq:etot}), such states can be further subdivided  into
  \begin{description}
    \item[a. Trapped (``T'') states:] these are particles with $E <0$. 
    \item[b. Quasi-trapped  (``Q'') states:] these are particles with $E \geq 0$.
  \end{description}
\end{description}
These three categories of particles are characterized by the particle numbers $\nu_C$, $\nu_T$ and $\nu_Q$, respectively, with 
the number of atoms $N=\nu_T+\nu_T+\nu_Q$. Even for a fixed value $N$, the three contributions can vary with time and with the incidence conditions and the surface parameters. In other words, an analysis of the dynamics of $\nu_C$, $\nu_T$ and $\nu_Q$ should provide detailed information on the gas-surface interaction and on the specific system.

It is generally expected that 
the quasi-trapped states (Q states) dominate 
at large incident angles $\theta$ (with respect to the surface normal) at first. 
The parallel momentum $p^{\parallel}$ does not change much during a single reflection event. A particle accelerates towards the surface and gains a large normal kinetic energy 
of $E^{\perp} \sim 80$\,meV by passing the depth of the physisorption well $E_0$ (cf.~table~\ref{tab1}). 
As a result, it collides with the surface close to the surface normal when the parallel momentum $p^{\parallel}$ remains practically unchanged. 

If a particle remains localized, its final momentum after every subsequent reflection 
can be projected sufficiently close to the surface plane, and, hence, the parallel momentum can be strongly perturbed by scattering at the atoms in the upper surface layer. Therefore, it is important to analyze the temporal evolution of the T and Q states and their equilibration mechanism.

As an example, the classification of surface states for three dynamical trajectories is presented in figure~\ref{fig:nstates}. 
In addition to the temporal evolution of the total energy $E(t)$, 
the distance $z(t)$ to the surface is displayed  as a function of time.  
It allows us to uniquely identify reflection events. 
Notice that a relatively fast energy exchange with the surface takes place at each reflection. Between the reflections the total energy is nearly conserved and correlated with the increase of the height $z(t)$. Once a particle moves in the opposite direction to the surface, it can be reflected back at the turning point specified by the kinetic and potential energy at the right boundary of the physisorption well. 
As our MD simulations show, here the total energy is conserved and, hence, the reflection can be treated as an elastic process. 

\subsection{Transitions between T, Q and C states}\label{ss:transitions}
Furthermore, the transition between the states along a particle trajectory is indicated by the 
line $n(t)$ in figure~\ref{fig:nstates}. The three values $n=\{-1,0,1\}$ are used to identify the three states $\{T,Q,C\}$, respectively.

The continuum states ($n(t)=1$) are observed before the first collision ($t/t_0\leq 0.4$) and for the final states 
with $(E_{\text{k}}^{\perp} + V) >0$,  where both energies are evaluated at the atom position $\vec r^\star$. 
The corresponding trajectories have no turning point and leave the surface region.
In between, such trajectories can experience multiple bounces and transitions between the trapped ($n(t)=-1$) and quasi-trapped ($n(t)=0$) states 
depending on the sign of the total energy $E(t)$. 

By analyzing a statistical ensemble of the trajectories of $n(t)$, 
a first ``physically'' relevant observation can be made. Before a particle is desorbed, it typically gets excited to a quasi-trapped state. The direct excitation probability from a trapped to a continuum state is significantly reduced at low lattice temperatures (e.g.\ $T_s=80$\,K), 
but it steadily increases with $T_s$, as will be shown in detail in section~\ref{rateSec}.

For large statistical ensembles of the trajectory states $n(t)$ 
we can explicitly evaluate the three fractions $N_j(t)=\nu_j(t)/N$ 
with $j=T$, $Q$, and $C$, where $\nu_j(t)$ is the number of trajectories with $n(t)=j$ at a given  instant of time. 
They satisfy the normalization condition $N_Q(t)+N_T(t)+N_C(t)=1$.
  
 \begin{figure}
 \begin{center} 
  \hspace{-0.cm}\includegraphics[width=0.5\textwidth]{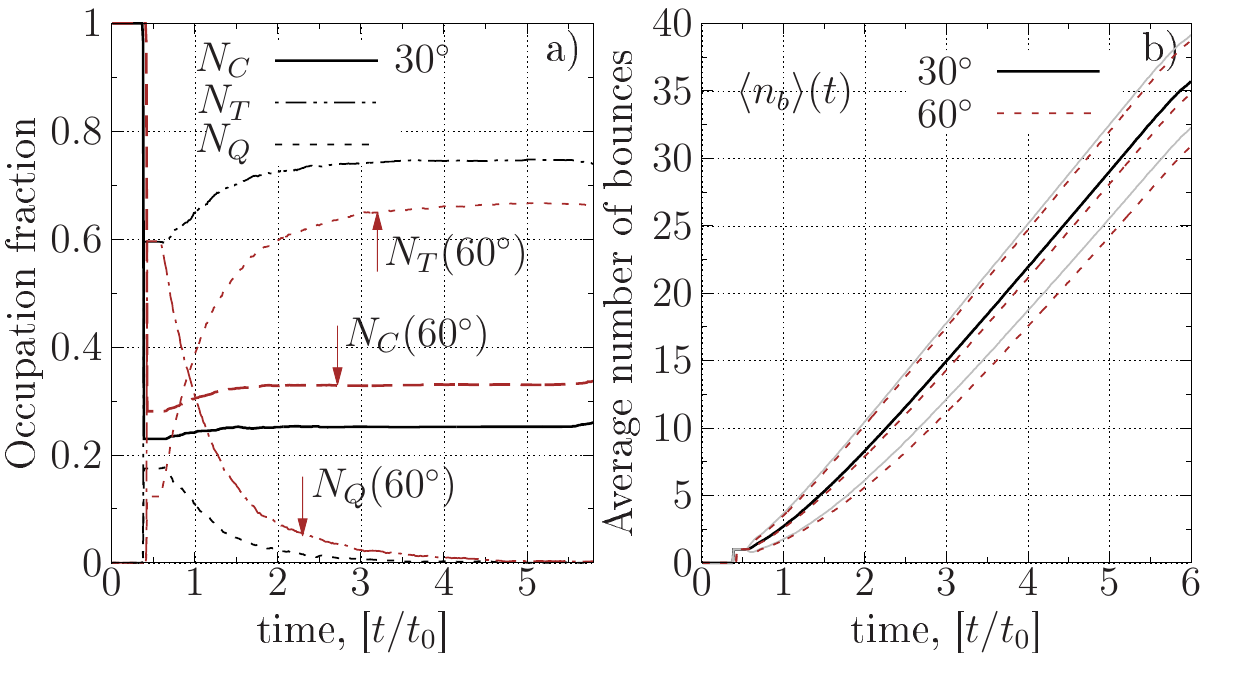} 
  \end{center}
  \vspace{-.80cm}
  \caption{\textbf{a)} Time dependence of the continuum ($N_C$), trapped ($N_T$) and quasi-trapped ($N_{Q}$) fractions of Ar atoms impacting a Pt(111) surface. The atoms have an energy of $E_i=15.93$\,meV for the incident angle $\theta=30^\circ$ and of $E_i=36.34$\,\text{meV} for $\theta=60^\circ$, respectively. 
  The lattice temperature equals $T_s=80$\,K. \textbf{b)} Average number of ``bounces'', $\avr{n_b}(t)$ (full black line). Pairs of  solid (dashed) lines correspond to the upper/lower bounds given by the variance: $\avr{n_b}\pm \sigma_{n_b}$.}
  \label{fig:ntypeT26ag3060}
  \end{figure}

Figure~\ref{fig:ntypeT26ag3060} shows the temporal evolution of these three fractions  of Ar atoms influencing a Pt(111) surface at $T_s=80$\,K 
for two conditions of incidence as well as the corresponding 
average number of bounces and their variance, $\sigma_{n_b}$, on the same time scale. 
The value $R_{st}(t)=N_{Q}(t)+N_T(t)=1-N_C(t)$ taken at $0.4\leq t/t_0\leq 0.6$ defines the initial sticking probability. 
Detailed results on the present system can be found in paper~II~\cite{paper2}. 
This quantity remains unchanged until a second bounce takes place. 
On average, this second bound occurs around $t\sim 0.7 t_0$. 

At low lattice temperature and incident atom energy, the initial sticking fraction 
remains quite large: $R_{st}\approx 0.75(0.65)$ for $\theta=30(60)^{\circ}$. 
The continuum states $N_C(t)$ show a saturation already for $t \geq 1.5 t_0$ corresponding to $9.8$\,ps. 
Within this period the trajectories experience  on average $n_b\sim 5$ bounces with the surface.  

In contrast, the ``pure'' trapped states show a 
convergent behavior 
at much later times, mainly due to an exponential-like decay of the quasi-trapped states. Within the time interval $0.4 < t/t_0 < 2.5$ there is a fast drop of $N_Q(t)$ from $58\%$ to $5\%$ for $\theta=60^\circ$ due to a fast conversion into trapped states. This can be seen from the increase in $N_T(t)$, which is accompanied by a practically constant value of $N_C(t)$ 
confirming that the particles remain localized near the surface.  

The case $\theta=30^\circ$ shows a similar trend, where only the absolute values $N_i(t)$ are different. 
The  fraction of trapped states is large: $N_T\sim 60\%$ already after the first reflection. 
This is to be expected, because  less energy is carried by the parallel component at smaller angles, i.e., $E_\mathrm{k}^{\parallel} \ll E_\mathrm{k}^{\perp}$. 
Hence, once the trapping condition 
$E_\mathrm{k}^{\perp}+V < 0$ 
is satisfied by the normal component, the total energy 
is negative 
in most cases ($E_\mathrm{k}+V< 0$). 
For $\theta=30^\circ$ the simulations predict that the initial fraction of quasi-trapped states can reach only $N_Q\sim 18\%$ and 
$N_Q$ completely decays into  trapped states  within $t\sim 3 t_0$. 

We conclude from figure~\ref{fig:ntypeT26ag3060} that the trapping becomes more efficient at smaller angles. 
At the lattice temperature $T_s=80$\,K the $N_Q$ fraction vanishes after $t\sim 5 t_0$ (33\,ps), while the $N_{T}$ fraction 
remains quasi-stationary with no noticeable thermal desorption observed on the simulated time scale $t< 6 t_0$ (40\,ps).

\section{Derivation of rate equations from kinetic theory}\label{s:derivation}

In this section we give a brief derivation of the time-dependent integral equations for the population of surface states. The transition probabilities can be expressed via microscopic quantities. Finally, we demonstrate how the kinetic equations can be reduced to a simplified description in terms of a rate equation model and  
energy-distribution-averaged transition coefficients. 
As we show in section~\ref{rateSec}, 
such a model allows for an efficient description of the time-resolved transitions between the trapped and continuum states. It also makes it possible to analyze the time, energy and temperature dependence of the sticking probability and the desorption rates.

Following the derivation presented by Brenig \cite{brenig_zphysb82}, we start with a general definition of the quantum mechanical transition probability between two states of the scattering atom. The transition of an atom from an initial (``i'')  state at time $t$ to a final (``f'') state at time $t+\tau$ is given in terms of the transition matrix element of the quantum mechanical time evolution operator $T_{f\mu,i\nu}=\langle f \mu | \hat T(\tau) | i \nu \rangle$. Here, the two indices  denote the states of an adsorbate atom, $\{i,f\}$, and of the substrate, $\{\mu,\nu\}$. 
Due to the unitarity of the evolution operator the transition probability obeys the general properties
\numparts
\begin{eqnarray}
&&\sum\limits_{f,\mu} \abs{\langle f \mu | \hat T(\tau) | i \nu \rangle}^2=1,\\
&&\sum\limits_{i,\nu} \abs{\langle f \mu | \hat T(\tau) | i \nu \rangle}^2=1.
\end{eqnarray}
\endnumparts

Starting from the initial state,
the transition probability to a final state can be defined as
\begin{eqnarray}
\abs{\langle f \mu | \hat T(\tau) | i \nu \rangle}^2=\abs{T_{f\mu,i\nu}}^2 \delta(\epsilon_f+E_{\mu}-\epsilon_i-E_{\nu}) \, \tau
\label{Tfi}
\end{eqnarray}
which depends linearly on $\tau$ and contains a delta function to guarantee the energy conservation. While this probability has the familiar form of Fermi's golden rule, the underlying assumptions should be recalled and critically assessed. First, it is assumed that their are no correlations between the adsorbate atom and the surface. This allows to write the total kinetic energy in the initial and final states as a sum of adsorbate kinetic energy and surface energy. Second, the linear dependence 
of the  transition probability
on $\tau$ is a  result 
of the 
perturbation theory 
and is valid for small time differences $\tau$. Third, the fact that the probability depends only on $\tau$ (and is independent of $t$) implicitly assumes that the system is stationary. Finally, the appearance of the delta function is a consequence of the Markov approximation which assumes that the correlation time has passed, and the energy spectrum has become stationary, see e.g. \cite{bonitz_qkt, bonitz_pla_96}. All these assumptions are justified in case of a macroscopically stationary surface which is only weakly perturbed by the scattering 
process of 
the adsorbate atom. 

In particular, we can safely assume 
for collisions of atoms with thermal and subthermal energy 
that the initial and final states of the substrate remain close to thermal equilibrium. The recoil energy, initially transferred to the surface layer, is rapidly dissipated to the bulk due to fast atomic vibrations 
taking place with the Debye frequency and a strong coupling between the substrate atoms 
when compared to the coupling with the adsorbate atoms. As a result, a saturation to the thermal equilibrium within the substrate is expected on  time scales much shorter than the adsorbate thermal accommodation time. 

Hence, as a further simplification,  we can introduce the transition probabilities averaged over the initial states of the substrate (specified by the Boltzmann factor $\rho_\nu$) 
\begin{eqnarray}
\avr{\abs{T_{fi}}}=\sum\limits_{\mu,\nu} \abs{\langle f \mu | \hat T(\tau) | i \nu \rangle}^2 \rho_\nu.
\end{eqnarray}
This allows us to obtain the 
temporal 
evolution of the system of gas atoms alone, which is obtained with the Pauli-Ansatz \cite{brenig_zphysb82}
\begin{eqnarray}
 n_f(t+\tau)=\sum\limits_i \avr{\abs{T_{fi}}} n_i(t),
\end{eqnarray}
which assumes low gas atom density so that all scattering events can be treated as independent of 
each other.
Using the definition of the kinetic coefficients 
according to $\bar T_{fi}=\avr{\abs{T_{fi}}}/\tau$, 
we end up with the system of coupled kinetic equations between the initial and final states 
\begin{eqnarray}
 \dot{n}_f(t)=\sum\limits_i \bar T_{fi}\, n_i(t) \, ,
 \label{rate1}
\end{eqnarray}
where the limit $\tau \to 0$ has been taken. 
This result is well known and directly expresses the connection between microscopic calculations~\cite{mc1,mc2} and kinetic theory~\cite{k1,k2}.

As a next step, we explicitly specify the quantum numbers $\{f,i\}$ of the gas atoms moving near the surface. 
Since 
quantum diffraction effects can be neglected due to the large mass of the gas atoms, 
a quasi-classical treatment can be used. 
Thus, it is sufficient to specify the states by the initial and final momenta   $\{\vec{p}_i,\vec{p}_f\}$, which
are additionally split into a tangential and parallel component relative to the surface. 
Consequently, we introduce the following notations
\begin{eqnarray}
 &&n_i(t)=n(p_i^\perp,\vec{p}_i^\parallel,t), \quad n_f(t)=n(p_f^\perp,\vec{p}_f^\parallel,t), 
  \label{nifstates}
\\
 &&\bar T_{fi}=\bar T(p_f^\perp,\vec{p}_f^\parallel,p_i^\perp,\vec{p}_i^\parallel) 
 \label{states}
\end{eqnarray}
in the kinetic equations.  
Note that the averaged transition  rate $\bar T$ 
(\ref{states}) depends solely on the incoming 
and outgoing 
momentum.

As a next step, we assume that the transition rate $\bar T_{fi}$ has only a weak directional dependence on the angle between the vectors $\vec{p}^{\parallel}_i$ and $\vec{p}^{\parallel}_f$. 
Its main dependence results from  two scalars, namely 
the perpendicular and  parallel kinetic energy components of the 
adsorbate atom, i.e., $\bar T_{fi}\approx \bar T_{fi}(\epsilon_f^\perp,\epsilon_f^\parallel,\epsilon_i^\perp,\epsilon_i^\parallel)$. 
This assumption is important for the derivation of the rate equations presented below. 
It 
can be justified by the theoretical treatment of a classical collision from vibrating surfaces~\cite{s1,s2,s3}. 

In particular, in Refs.~\cite{s4,s5} an explicit expression for the 
zeroth-order reflection coefficient 
$R^0$ 
has been derived for an atomic projectile colliding with a surface consisting of discrete scattering centers (of mass $M$) whose initial momenta are given by an equilibrium distribution at the lattice temperature $T_s$. This expression is given by
\begin{eqnarray}
 \frac{\db R^0(\vec p_f,\vec p_i)}{\db E_f \db \Omega_f}=&&\frac{m^2 \abs{\vec p_f}}{8\pi^3 \hbar^4 p_i^\perp} \abs{\tau_{fi}}^2 \left( \frac{\pi}{k_B T_s \Delta E_r} \right)^{1/2} \nonumber \\ &&\times \exp \left(- \frac{(E_f-E_i+\Delta E_r)^2}{4k_B T_s \Delta E_r} \right)\,,
\end{eqnarray}
and shows a dependence on several key parameters. 
Here, $p_i^\perp$ is the $z$-component of the incident momentum, $\abs{\tau_{fi}}^2$ is the form factor of the scattering center, which depends on the interaction potential, and $E_{f(i)}$ is the kinetic energy after (before) the collision. The only parameter, which contains a directional dependence, is the recoil energy expressed as 
\begin{eqnarray}
\Delta E_r=\frac{\Delta\vec p_{fi}^2}{2M}
=\frac{(p_{f}^{2\perp}-p_{i}^{2\perp})}{2M}+ \frac{(p_{f}^{2\parallel}-p_{i}^{2\parallel})}{2M}-\frac{\vec p_f^\parallel \vec p_i^\parallel}{M}.
\nonumber
\end{eqnarray}
Due to the inelastic and Brownian-like character of the scattering processes, the contribution of the first two terms should dominate. The directional dependence 
(third term) is expected to be weak. Thus, we can use $\Delta E_r\approx \Delta E_r(\epsilon_f^\perp,\epsilon_f^\parallel,\epsilon_i^\perp,\epsilon_i^\parallel)$. Finally, it is assumed that the scattering amplitude $\abs{\tau_{fi}}$ is a constant, with a value derived for hard sphere scattering.

Now we can proceed and explicitly define the population of surface states and the inter-state transition probabilities 
by the dependence on the kinetic energy components
\begin{eqnarray}
 n_i(t)&=&n(\epsilon_i^\perp,\epsilon_i^\parallel,t), \quad n_f(t)=n(\epsilon_f^\perp,\epsilon_f^\parallel,t),\\
 \bar T_{fi} &=& \bar T(\epsilon_f^\perp,\epsilon_f^\parallel,\epsilon_i^\perp,\epsilon_i^\parallel).
 \label{states2}
\end{eqnarray}
In the following we omit the subscripts ``i'' and ``f'' and indicate the initial and final states, instead, by using different energy symbols: 
\begin{eqnarray}
\epsilon &=\{\epsilon_i^{\perp},\epsilon_i^\parallel\}\,,
\label{eq:e}\\
\epsilon^\prime &=\{\epsilon_f^{\perp},\epsilon_f^{\parallel}\}\,.
\label{eq:eprime}
\end{eqnarray}

Following the detailed discussion given in section~\ref{model-states}, all possible surface and scattering states can be classified into three categories using  
energy criteria. Now, the trapped, quasi-trapped and continuous particle fractions introduced in section~\ref{model-states} can be  explicitly defined via integration of the time-dependent energy distribution functions, which define the population of different surface states, 
according to 
\numparts
\begin{eqnarray}
 N_T(t)&=\int\limits_0^{\abs{V}} \db \epsilon_{\perp} \int\limits_{0}^{{V-\epsilon^{\perp}}} \db \epsilon^{\parallel} n(\epsilon^{\perp},\epsilon^{\parallel},t)=\int\limits_{\Omega_T} \db \epsilon \, n(\epsilon,t),\label{defstates1}\\
 N_Q(t)&=\int\limits_0^{\abs{V}} \db \epsilon^{\perp} \int\limits_{V-\epsilon^{\perp}}^{\infty} \db \epsilon^{\parallel} n(\epsilon^{\perp},\epsilon^{\parallel},t)=\int\limits_{\Omega_Q} \db \epsilon \, n(\epsilon,t),\\
 N_C(t)&=\int\limits_{\abs{V}}^{\infty} \db \epsilon^{\perp} \int\limits_{0}^{\infty} \db \epsilon^{\parallel} n(\epsilon^{\perp},\epsilon^{\parallel},t)=\int\limits_{\Omega_C} \db \epsilon \, n(\epsilon,t).
 \label{defstates3}
\end{eqnarray}
\endnumparts
Here,  we introduced the shorthand notation $\int_{\Omega_s} \db \epsilon$ with $s=$T, Q, C 
for the distinction of different states and the integration limits. 
This energy integral always comprises 
a double integration over the tangential and normal kinetic energy components. 
The upper integration limit for the tangential component specifies that the trapped and quasi-trapped states stay localized (bound) near the surface due to the condition $(-\abs{V}+\epsilon_{\perp})<0$ (or 
$0\leq \epsilon_{\perp} \leq \abs{V}$). Here, 
$\abs{V}=\abs{V(\vec r^\star)}$ is the potential energy in the physisorption well where the total particle energy is nearly conserved. 
The second integral, i.e., that over the parallel component, introduces a distinction between bound and continuous states.

Using the definitions (\ref{eq:e}) and (\ref{eq:eprime}) 
for the notation of the initial and final state energies, 
we can rewrite the kinetic equation~(\ref{rate1}) in the form
\begin{eqnarray}
 \dot{n}(\epsilon,t)=\int\limits_{0}^{\infty} \db \epsilon^\prime \, T(\epsilon,\epsilon^\prime) n(\epsilon^\prime,t) - \int\limits_{0}^{\infty} \db \epsilon^\prime \, T(\epsilon^\prime,\epsilon) n(\epsilon,t).
\nonumber
\end{eqnarray}
This equation specifies the temporal evolution of the three types of states introduced in Eqs.~(\ref{defstates1})-(\ref{defstates3}) by taking the time derivative. 
This yields  the result 
\begin{eqnarray}
\dot{N_s}(t)=\int\limits_{\Omega_s} \db \epsilon \int\limits_{0}^{\infty} \db \epsilon^\prime\,\left[ T(\epsilon,\epsilon^\prime) n(\epsilon^\prime,t)- T(\epsilon^\prime,\epsilon) n(\epsilon,t) \right] 
\label{Ns}
\end{eqnarray}
with $s$ = Q, T, C. The  first   term  defines  the incoming  flux  from  all  possible  states $\epsilon^\prime$ 
and the second term represent the  outgoing flux from the state $\epsilon$.  
 The inner integral over $\epsilon^\prime$ can be splitted 
 into the contribution of different surface states, $\int =\int_{\Omega_T} +\int_{\Omega_Q}+ \int_{\Omega_C}$. This allows to introduce the energy-resolved inter-state transition rates, which carry the energy-dependence of initial state ($s^{\prime}$ = Q, T, C) and are integrated over the energy of the final state ($s$ = Q, T, C) according to
\begin{eqnarray}
T_{ss^\prime}(\epsilon^{\prime})=\int_{\Omega_s} \db \epsilon \, T(\epsilon,\epsilon^\prime)\,,     
\nonumber
\end{eqnarray}
for $\epsilon^\prime \in \Omega_{s^\prime}$. The diagonal terms with $s=s^\prime$ are mutually cancelled in Eq.~(\ref{Ns}) and, therefore, they can be excluded from the consideration. We rewrite Eq.~(\ref{Ns}) to the form
\begin{eqnarray}
  \dot{N_s}(t) &= &\sum\limits_{s^\prime\neq s} \bigg[\, \int\limits_{\Omega_{s^\prime}}\db \epsilon^\prime \, T_{ss^\prime}(\epsilon^\prime) n(\epsilon^\prime,t) 
 \bigg. \nonumber  \\
 && 
 \qquad 
 \bigg.
 - \int\limits_{\Omega_s}\db \epsilon \, T_{s^\prime s}(\epsilon) n(\epsilon,t)\bigg] 
 \label{Tsss}
\end{eqnarray}
and end up with a set of rate equations 
\begin{eqnarray}
 \dot{N_s}(t)=\sum\limits_{s^\prime\neq s} \left[\bar T_{ss^\prime}(t) N_{s^\prime}(t)-\bar T_{s^\prime s}(t) N_{s}(t) \right] 
 \label{Ns2}
\end{eqnarray}
for the final states with $s$ = Q, T, C, 
where 
\begin{eqnarray}
 \bar{T}_{ss^\prime}(t)=\frac{\int_{\Omega_{s^\prime}}\db \epsilon^\prime \, T_{ss^\prime}(\epsilon^\prime) n(\epsilon^\prime,t)}{\int_{\Omega_{s^\prime}}\db \epsilon^\prime \, n(\epsilon^\prime,t)} 
 \label{Tss}
\end{eqnarray}
are the energy distribution-averaged transition coefficients.

This set of coupled equations can be further simplified, once the system reaches a quasi-equilibrium state. This regime can be identified from the convergence of the momentum and energy distribution functions to a quasi-stationary form, $n(\epsilon,t)=\gamma(t)\cdot n(\epsilon,t^E)$. The distribution converges to the shape $n(\epsilon,t^E)$ specified by the quasi-equilibration time $t^E$, and stays unchanged up to some time-dependent scaling factor $\gamma(t)$. 
In this regime the transition rates~(\ref{Tss}) also become  
time-independent with the constant values 
\begin{eqnarray}
 \bar{T}_{ss^\prime}^E=\frac{\int_{\Omega_{s^\prime}}\db \epsilon^\prime \, T_{ss^\prime}(\epsilon^\prime) n(\epsilon^\prime,t^E)}{\int_{\Omega_{s^\prime}}\db \epsilon^\prime \, n(\epsilon^\prime,t^E)} 
 \label{TssTe}
\end{eqnarray}
defined by the quasi-equilibrium energy distribution of the system in the different surface states (e.g.\ $s^{\prime}$= Q, T, C.). 
This result will be used in the next section, where we demonstrate how these specific values can be derived from MD simulations.

\section{Rate equation model}\label{rateSec}
Based on Eq.~(\ref{Ns2}), the temporal evolution 
of the three types of surface states can be analyzed 
on the quantitative level 
by the set of rate equations
\numparts
\begin{eqnarray}
 \dot{N}_{Q} &=& - (T_{TQ}+T_{CQ})N_{Q} +  T_{QT} N_{T},\label{eq1}\\
 \dot{N}_{T} &=& -  (T_{QT}+T_{CT}) N_{T} +  T_{TQ} N_{Q}, \label{REQeq2} \\
 \dot{N}_{C} &=& - (\dot{N}_{Q} + \dot{N}_{T})=T_{CT}N_{T}+ T_{CQ}N_{Q} \, ,
 \label{REQ}
\end{eqnarray}
\endnumparts
where the backwards transitions 
from the continuous to the bound states are assumed to be negligible. 
The fractions of atoms in the three states change with time due to different decay channels. 
For instance,  the first two terms in Eq.~(\ref{eq1}) 
take into account the decay of the Q states into 
T and C states with the transition rate $T_{TQ}$ and $T_{CQ}$, respectively. 
This means we use the notation $T_{\alpha \beta}$ for transitions of the type $\beta \to \alpha$. 
According to~(\ref{Tss}), 
the transition rates in the set of equations (\ref{eq1})-(\ref{REQ}) are generally time-dependent, i.e.,  $T_{\al\be}=T_{\al\be}(t)$. 
They crucially depend on the energy distribution function of the initial and final state  
in general. Hence,  the energy distribution function 
is non-stationary as well until the system reaches  quasi-equilibrium. 

   \begin{figure}
  \begin{center} 
  \hspace{-0.cm}\includegraphics[width=0.5\textwidth]{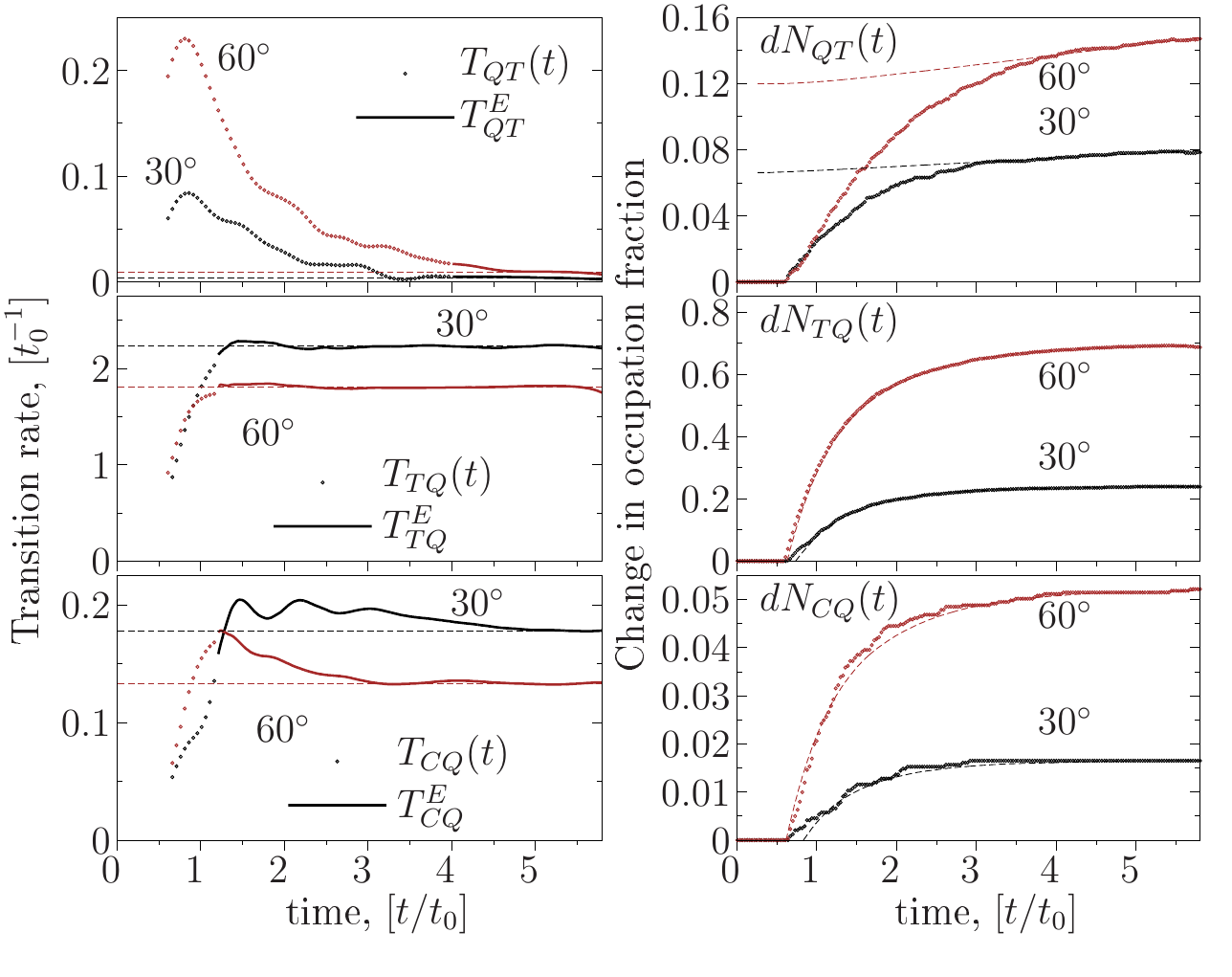} 
  \end{center}
  \vspace{-.80cm}
  \caption{\textbf{Left:} Time dependence of the transition rates $T_{\al\be}(t)$. The monoenergetic beam has the energy $E_i=15.93(36.34)$\,meV for the incident angle $\theta=30(60)^\circ$. Lattice temperature: $T_s=80$\,K. \textbf{Right:} Change in the  population $\db N_{\al\be}(t)$ of the three states ($\al=$Q, T, C) with the normalization $N_{Q}(t)+N_{T}(t)+N_{C}(t)=1$ 
  due to the decay channel $\be \rightarrow \al$. The dashed line, deviating from the MD data (dotted curve) at small times, is the prediction of the 
  rates at quasi-equilibrium 
  extrapolated to $t\leq t^E$.}
  \label{fig:rateT26ag3060}
  \end{figure} 

\subsection{Reconstruction of the transition rates from the MD simulations}\label{ss:reconstruction}
Now we demonstrate how the rates $T_{\al\be}(t)$ can be extracted from the 
MD 
simulation data. In figure~\ref{fig:rateT26ag3060} we plot the change in the  population of three states (Q, T and C) due to the net transition fluxes  
\numparts
\begin{eqnarray}
\db N_{QT}(t) &=T_{QT}(t)\, N_T(t) \, \db t,\\ 
\db N_{TQ}(t) &=  T_{TQ}(t)\, N_Q(t)\, \db t, \\
\db N_{CQ}(t) &= T_{CQ}(t)\, N_Q(t)\, \db t. 
\end{eqnarray}
\endnumparts
Two incident conditions 
are compared, which 
correspond to 
the energy  
$E_i=15.93$\,meV at the angle $\theta=30^\circ$ 
and 
$E_i=36.34$\,meV at  $\theta=60^\circ$, respectively, for the lattice temperature $T_s$ of 80\,K.  
Each of the decay channels can be uniquely identified by analyzing the initial and final state along the trajectory for every bounce event shown in figure~\ref{fig:nstates} for several examples. 
The transition rate can be extracted by performing the numerical differentiation 
\begin{eqnarray}
 T_{\al\be}(t)=\frac{\db N_{\al\be}(t)}{\db t}\frac{1}{N_{\be}(t)} 
 \label{Tba}
\end{eqnarray}
for the curves shown in figure~\ref{fig:rateT26ag3060}. 

The following procedure has been used. First, the MD data have been smoothed  by a Gaussian kernel
\begin{eqnarray}
 &&f_G(t)=\sum\limits_{j=1}^{N} \frac{G_j(t)}{G(t)}\, f(t_j),\nonumber\\
 &&G_j(t)=e^{-\frac{(t-t_j)^2}{2h^2}},\quad 
  G(t)= \sum\limits_{j=1}^{N} G_j(t),
 \label{gauss}
\end{eqnarray}
where $h=\nu\cdot \Delta t$ ($\nu=2\ldots 10$), and $N$ is the total number of points on the curve. The new values are evaluated using the weighted contribution on 
neighboring points, and, hence, the statistical fluctuations at each point $t_j$ are suppressed. The  differentiation of  $f_G$ with respect to time  leads to 
\begin{eqnarray}
\frac{\db}{\db t} f_G(t)=-\sum\limits_{j=1}^{N} \frac{G_j(t)}{G(t)} \frac{t-t_j}{h^2} [f(t_j)-f_G(t)].
\end{eqnarray}
The smoothness of the derivative can be controlled by the parameter $h$ and adjusted to give 
better agreement with the MD data. Typically, the value $\nu\sim 6$ was found to be a reasonable choice.

Alternatively, the transition rates can be expressed via the integral form
\begin{eqnarray}
 \db N_{\al\be}(t)\big|_{t> t^{E}} &=& \int\limits_0^{t^E} T_{\al\be}(\tau) N_\be(\tau) \db \tau+ T_{\al\be}^E \int\limits_{t^E}^{t} N_\be(\tau) \db \tau,\nonumber\\
 &=& \db N_{\al\be}(0,t^E)+T_{\al\be}^E \cdot \db N_\be(t^E,t)\label{Teq}.
\end{eqnarray}
Here, we used the assumption that  the system state $\be$ has reached a quasi-equilibrium state 
for $t\geq t^{E}$. Then, the transition rate has only a weak time-dependence, i.e.,  $T_{\al\be}(\tau)|_{\tau\geq t^{E}} \approx T_{\al\be}^{E}$. 
For the example shown in figure~\ref{fig:ntypeT26ag3060}, the equilibration time $t^E$ 
is about $3t_0$. 
Finally, the quasi-equilibrium transition rate can be determined from Eq.~(\ref{Teq}) as the ratio of the integrated population of states
\begin{eqnarray}
 T_{\al\be}^{E}(t)=\frac{\db N_{\al\be}(t)-\db N_{\al\be}(0,t^E)}{\db N_\be(t^E,t)}.
 \label{TbaE}
\end{eqnarray}
By a proper choice of the equilibration time $t^E$, 
the estimated rate $T_{\al\be}^{E}(t)$ should exhibit only a weak time dependence and represent the asymptotic limit of the more general time-dependent rate in Eq.~(\ref{Tba}).

The  data 
$\{\db N_{\al\be},\,\db N_\be\}$ 
are provided by the MD simulations. To determine the transition rates and the population of states more accurately, we have used the statistical averages over several thousand trajectories. 
We performed the analysis similar to figure~\ref{fig:nstates} 
 for each set of initial parameters ($T_s,E_i,\theta$) 
and determined the fluxes between different states.

\subsection{Test of the approach}\label{ss:test}
The comparison of the transition rates  given by Eqs.~(\ref{Tba}) and (\ref{TbaE}) is demonstrated in figure~\ref{fig:rateT26ag3060}. 
We choose the time of equilibration 
$t^E=4t_0$ for the rate $T_{QT}$ and  $t^E=1.2 t_0$ for $T_{TQ},T_{CQ}$. 
The results of Eq.~(\ref{Tba}) are represented by the dots, and those from Eq.~(\ref{TbaE}) by the solid lines. Both should match at $t=t^E$. The choice of $t^E$ in each case needs some adjustments, such that $T^E_{\al\be}$ should be nearly constant for $t> t^E$. The quality of the quasi-equilibrium approximation 
can be checked on the right panel. The rates $T_{\al\be}(t)$, reconstructed from Eq.~(\ref{Tba}), accurately fit the MD data (presented by the dotted curves) for all simulation times and the changes in state populations ($\db N_{QT}$, $\db N_{TQ}$ and $\db N_{CQ}$). 

Due to a statistical noise present in the MD data, the extracted rates exhibit artificial oscillations, see e.g. $T_{CQ}(t)$. At longer times they can be successfully removed using the 
equilibrium estimator Eq.~(\ref{TbaE}), 
which is much less influenced by the statistical noise. The corresponding 
equilibrium rate
$T^E_{\al\be}$ is shown by the horizontal dashed curves on the left panels. 
In particular, 
the rate $T^E_{QT}$ 
for the incident angle $\theta=30^\circ$($60^\circ$)  
can reproduce the change in the population of quasi-trapped states, $\db N_{QT}(t)$, for $t>3t_0$ ($t>4t_0$). The left panel in figure~\ref{fig:rateT26ag3060} shows that the corresponding non-equilibrium rate $T_{QT}(t)$ actually  converges slowly to $T_{QT}^E$ on a time scale, which depends on the incident angle.

A similar analysis can be performed for $T_{TQ}^E$ and $T_{CQ}^E$. Here, we observe that the 
quasi-equilirium assumption 
can be introduced at a significantly earlier moment. Using $t^E=1.2t_0$ we can nearly exactly reproduce the MD data (see the right panel) for the conversion of the quasi-trapped states to the trapped state, i.e., $\db N_{TQ}(t)$, and with some deviations observed at $t<2t_0$ the desorption of the quasi-trapped states to the continuum, i.e., $\db N_{CQ}(t)$. The rates $T_{TQ}(t)$ and $T_{CQ}(t)$ stay practically constant over the entire simulation period, starting at $t \sim t_0$. However, at $t< t_0$ the rates drop to smaller values. The later behaviour is artificial and should be explained by the smoothing procedure applied in Eq.~(\ref{gauss}). It perturbs the slope of the MD curve around the time of second reflection from the surface ($t\sim 0.6t_0$).

In summary, we have demonstrated the efficiency of the rate equation model and provided the proof of convergence to quasi-equilibrium after some transient time $t^E$.

\begin{figure}
  \begin{center} 
  \hspace{-0.cm}\includegraphics[width=0.5\textwidth]{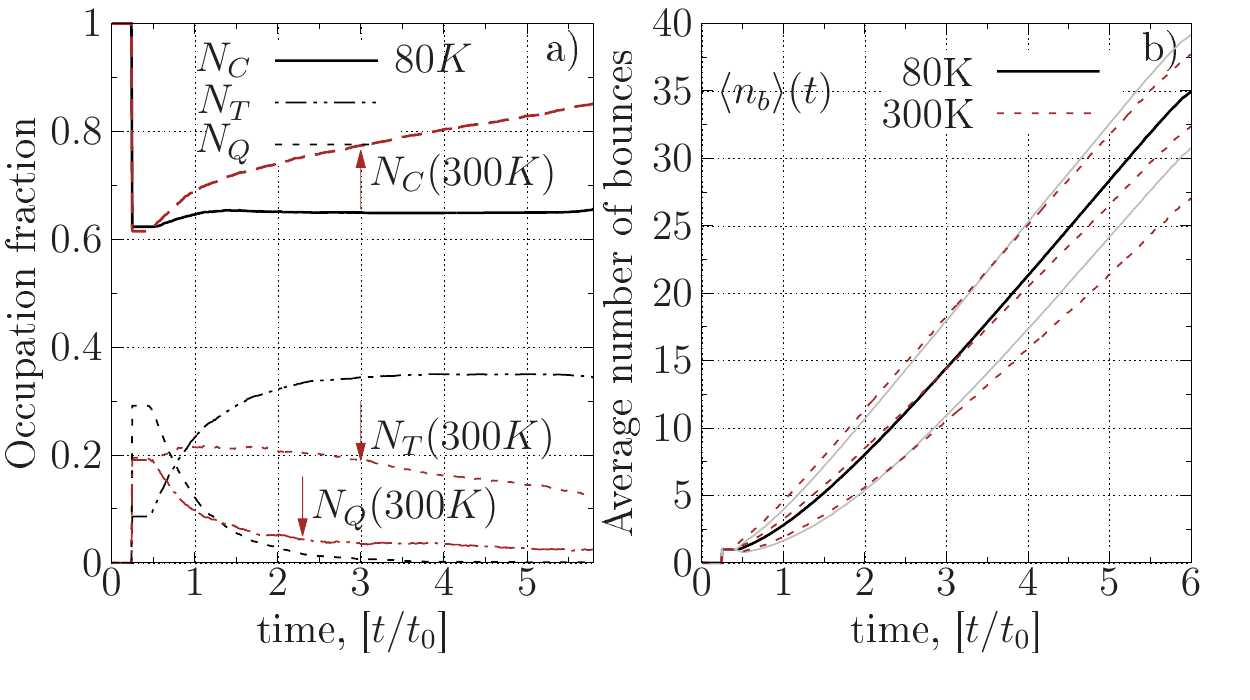} 
  \end{center}
  \vspace{-.80cm}
  \caption{\textbf{a)} Time dependence of the continuum ($N_C$), trapped ($N_T$) and quasi-trapped ($N_{Q}$) fractions for Pt(111). The monoenergetic atoms have an energy of $E_i=49.42\text{meV}$ for the incident angle $\theta=30^\circ$. The lattice temperatures are $T_s=80$ and $300$\,K. \textbf{b)} Average number of bounces $\avr{n_b}(t)$. The two dotted lines correspond to the upper/lower bounds given by the variance: $\avr{n_b}\pm \sigma_{n_b}$.}
  \label{fig:ntypeT26T10ag30e20}
  \end{figure} 

\begin{figure}
  \begin{center} 
  \hspace{-0.cm}\includegraphics[width=0.5\textwidth]{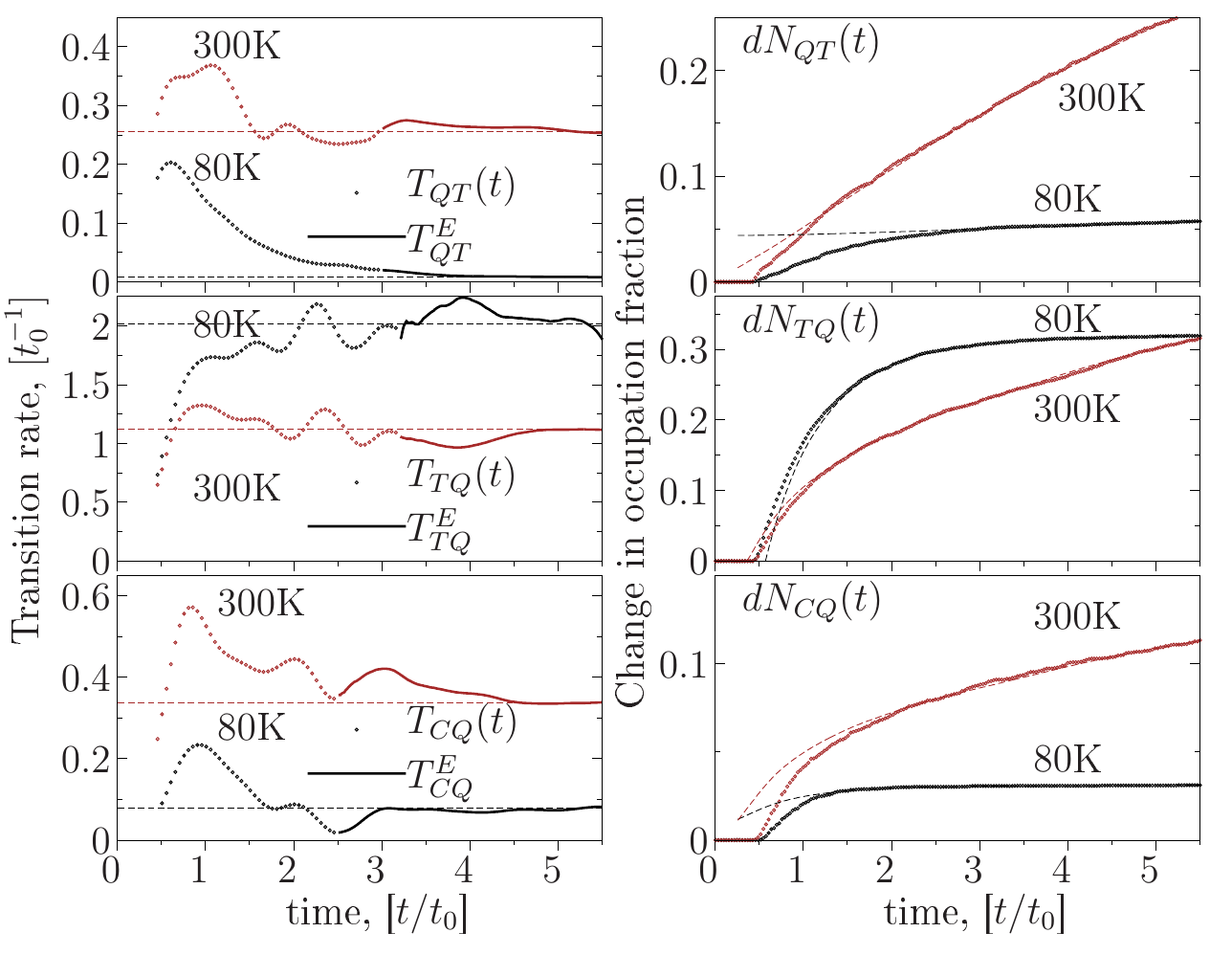} 
  \end{center}
  \vspace{-.80cm}
  \caption{\textbf{Left:} Time dependence of the transition rates $T_{\al\be}(t)$ for the 
  conditions in figure~\ref{fig:ntypeT26T10ag30e20}. 
  \textbf{Right:}
Change in the population, $\db N_{\al\be}(t)$, for$T_s=80$ and $300$\,K.}
  \label{fig:ratesT26T10ag30}
  \end{figure}

The next important point 
concerns the analysis of the impact of the lattice temperature $T_s$. 
In figures~\ref{fig:ntypeT26T10ag30e20} and \ref{fig:ratesT26T10ag30} we compare 
results for the two temperatures $T_s=80$\,K and $300$\,K. 
The incident angle is $\theta=30^\circ$ and the initial gas energy is $E_i=49.42$\,meV. 
Compared to the case with $E_i=15.93$\,meV shown in figure~\ref{fig:ntypeT26ag3060}, 
we observe an increase by a  factor of $3$  in $N_C(t)$ at $t\sim 0.4t_0$ 
in figure~\ref{fig:ntypeT26T10ag30e20} for $T_s=80$\,K, 
i.e., a much higher fraction of particles is reflected  after the first bounce. 
At the same time, 
the initial population of the trapped states is reduced by factor 6 (from $0.6$ to $0.09$), while the population of Q states is increased by $50\%$ (from $0.18$ to $0.29$). 
During $t\leq 4t_0$ the Q states practically vanish due to the decay channel Q $\rightarrow$ T. 
Furthermore,  the trapped state $N_T(t)$ demonstrate stability against the thermal desorption 
at this lattice temperature of 80\,K. 
However, the temporal evolution of the state populations is very similar to the case with the incident energy $E_i=15.93$\,meV (figure~\ref{fig:ntypeT26ag3060}).

The situation changes at the higher lattice temperature $T_s=300$\,K shown in figure~\ref{fig:ntypeT26T10ag30e20}. 
The initial reflection coefficient $N_C(t)|_{t \sim 0.4t_0}$ is similar to the low-temperature case ($T_s=80$\,K), but then  
it rapidly increases with two characteristic rates. 
As shown in  figure~\ref{fig:ratesT26T10ag30}, 
a higher rate is found for $0.4\leq t/t_0\leq 1.2$ due to a fast decay of the Q states 
into the C and T states: 
$T_{CQ} \geq 3.5$ and $T_{TQ} \geq 9$. 
Because $T_{TQ}$ is larger than $T_{CQ}$, 
the conversion to the trapped states is the dominant process when the fraction $N_{Q}$(t) is large.

This trend changes at $t\sim t_0$ (or after 2-3 bounces with the surface). 
As it is becomes clear from figure~\ref{fig:ntypeT26T10ag30e20}
the $N_Q$ fraction is reduced by a factor of 2, 
 while the $N_T$ fraction
reaches  a  local  maximum.   For $t\sim t_0$ 
the  trapped  states
dominate and are steadily converted to the continuum states.
Note that the decay of the T states takes place much faster than
the decay of the Q states.

 The $N_Q$ fraction first saturates around $4\%$ and then slowly decays to $2\%$ at $t\approx 6t_0$. This behavior can be explained on basis of the rate equations. As shown in  figure~\ref{fig:ratesT26T10ag30} for $T_s=300$\,K, 
 the rates of the mutual conversion 
 T $\leftrightarrow$ Q between the states T and Q 
 differ by a factor of 4. However, they contribute in the rate equations~(\ref{eq1}) and 
 (\ref{REQeq2}) 
 being multiplied by the population factors 
 and, hence, all incoming and outgoing net fluxes can be compensated for a given state (e.g. $\dot{N}_Q(t)\approx 0$) if the detailed balance condition 
\begin{eqnarray}
 T_{QT}(t) \, N_T(t)\approx [T_{CQ}(t)+T_{TQ}(t)]\, N_Q(t) 
 \label{dbc}
\end{eqnarray}
approximately holds. 
Indeed, this condition can be satisfied for the case shown in figure~\ref{fig:ntypeT26T10ag30e20} 
when the ratio of both populations increases to $N_T(t)/ N_Q(t)\geq 4$  for $t>2t_0$. The proof why the relation~(\ref{dbc}) should hold in general will be given in section~\ref{ss:analytics}.

One can use the rates shown in figure~\ref{fig:ratesT26T10ag30} 
to analyze how fast the system reaches quasi-equilibrium. Using the data for  $T_{QT}(t)$ and $T_{TQ}(t)$,  we estimate $t^E \approx 1.5t_0$ for $T_s=300$\,K and $t^E \approx 3 t_0$ for $T_s=80$\,K. Hence, 
a higher lattice temperature favors a faster adsorbate equilibration. A more quantitative discussion of the convergence to  quasi-equilibrium is presented in 
paper~II~\cite{paper2}.
\begin{figure}
  \begin{center} 
  \hspace{-0.cm}\includegraphics[width=0.51\textwidth]{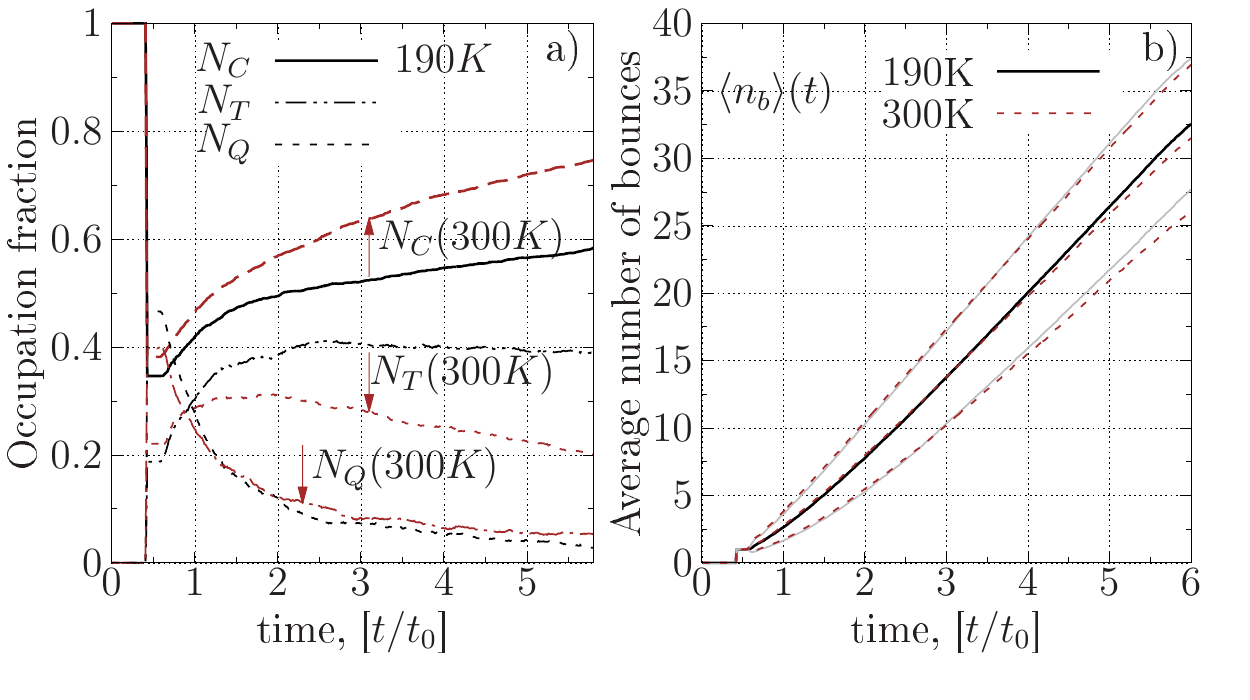} 
  \end{center}
  \vspace{-.80cm}
  \caption{\textbf{a)} Time dependence of the continuum ($N_C$), trapped ($N_T$) and quasi-trapped ($N_{Q}$) fractions for Pt(111). The monoenergetic beam has the energy $E_i=36.34\,\text{meV}$ and the incident angle $\theta=60^\circ$. Two lattice temperatures are compared: $T_s=190$\,K and $300$\,K. \textbf{b)} Average number of bounces $\avr{n_b}(t)$. The pairs of dotted lines correspond to the upper and lower bounds given by the variance: $\avr{n_b}\pm \sigma_{n_b}$.}
  \label{fig:ntypeT06ag60}
  \end{figure}
\begin{figure}
  \begin{center} 
  \hspace{-0.cm}\includegraphics[width=0.5\textwidth]{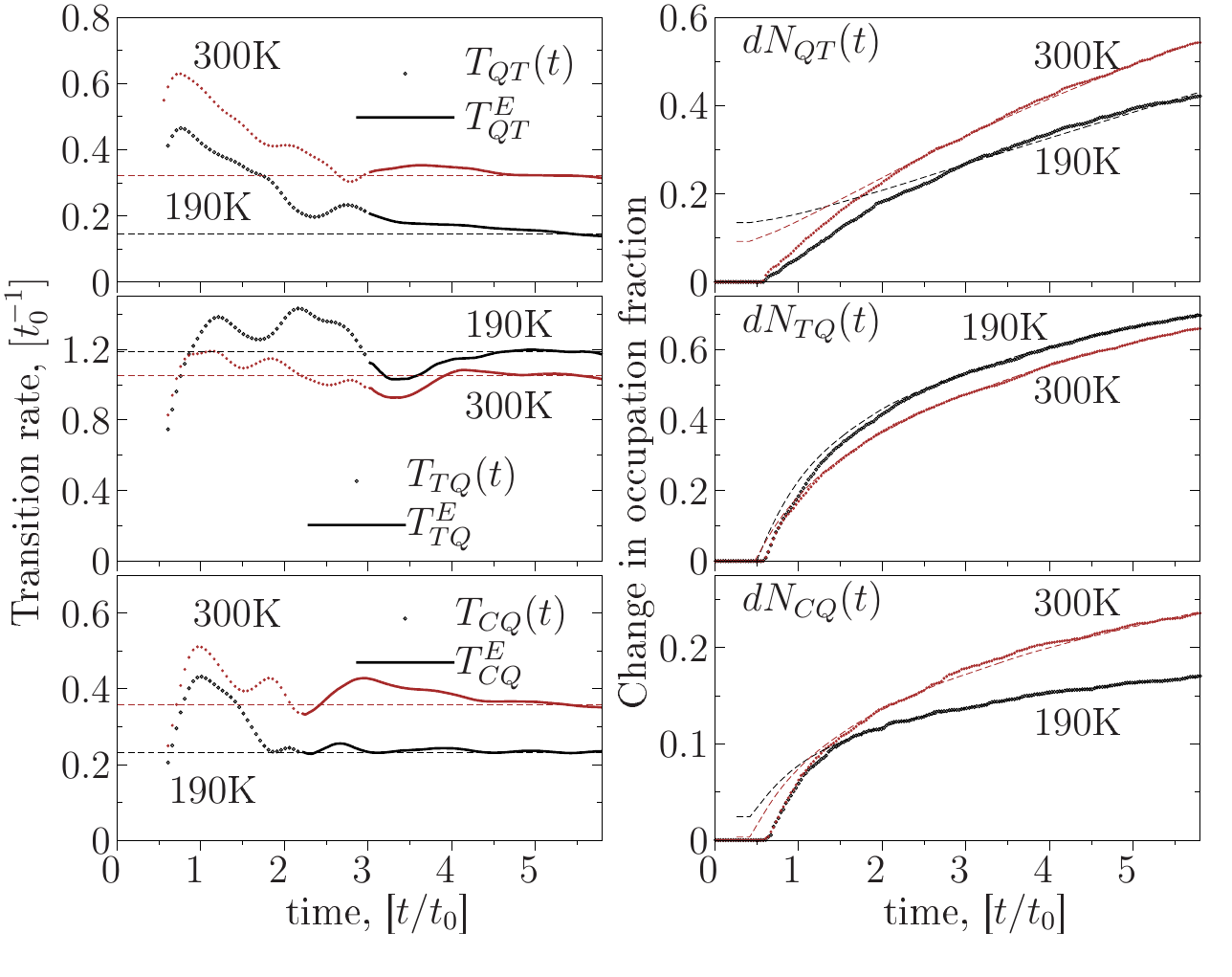} 
  \end{center}
  \vspace{-.80cm}
  \caption{\textbf{Left:} Time dependence of the transition rates $T_{\al\be}(t)$ for the case shown in Fig.~\ref{fig:ntypeT06ag60}. Lattice temperatures: $T_s=190$K and $300$K.
   \textbf{Right:} Change in the population $\db N_{\al\be}(t)$ for both temperatures.}
  \label{fig:ratesT06ag60}
  \end{figure} 
  
The corresponding analysis of 
the temperature effects for the larger incident angle $\theta=60^{\circ}$ 
is shown in figures~\ref{fig:ntypeT06ag60} and \ref{fig:ratesT06ag60} 
for the two temperatures $T_s=190$ and $300$\,K. 
Due to the larger incident angle, the initial population of quasi-trapped states is above $40\%$  and slightly decreases with increasing temperature. The population of trapped states is 
around $20\%$ for both the surface temperatures. 
While the values specified at the time of the first bounce ($t\leq 0.5t_0$) are similar, 
their temporal evolution  is quite different due to an enhanced thermal desorption at 
the larger temperature $T_s=300$\,K. 
This can be clearly resolved from the temporal evolution of the continuum states, 
given by the curves $N_C(t)$. 
The trapped fraction $N_T(t)$ seems to saturate 
during  the time-interval $1.5\leq t/t_0\leq 2$, and it 
subsequently decreases for $t> 2t_0$ due to thermal desorption going faster at $T_s=300$\,K.  
This can also be identified by a faster increase of the continuum fraction $N_C(t)$. 
During the period $2\leq t/t_0\leq 6$ the $N_T(t)$ fraction is reduced by a factor 
of $1.5$ from $30$ to $20\%$, 
while it is reduced only by $2\%$ at $T_s=190$\,K. 
 
In contrast, the $Q$ states show only a weak dependence on $T_s$. 
They decay with a similar slope for both lattice temperatures. 
As in figure~\ref{fig:ntypeT26T10ag30e20}, we  observe a fast conversion via the two channels Q $\rightarrow$ T and Q $\rightarrow$ C at first. 
The first channel dominates due to a high rate $T_{TQ}$ (cf.~figure.~\ref{fig:ratesT06ag60}). 
Its value  decreases only slightly with increasing temperature, 
whereas the asymptotic value of the two other rates $T^E_{QT}$ and $T^E_{CQ}$ 
at quasi-equilibrium drops by a factor of 2 by lowering the temperature from $T_s=300$ to $190$\,K.  This finding seems quite reasonable. 
The de-excitation transition takes place, when an excited state (a quasi-trapped trajectory) releases an energy to go into a lower energy state (a trapped trajectory). 

In contrast, the two 
transitions T $\rightarrow$ Q and Q $\rightarrow$ C  require that a finite portion of energy 
should be supplied from the lattice. In this case, the excitation probability
should scale with the population of phonon modes and depend on  multi-phonon excitations. 
Here, a strong temperature dependence is expected. Indeed, the data 
for  $T_{QT}$ and $T_{CQ}$ presented in figures~\ref{fig:ratesT26T10ag30} and~\ref{fig:ratesT06ag60}    confirm this expectation and demonstrate a clear temperature dependence. 
However, if the rates are compared at the same lattice temperature, but different incident angles 
(see $\theta=30^{\circ}$ and $60^{\circ}$ in figure~\ref{fig:rateT26ag3060}), the difference in the rates does not exceed $10\%$.

We can conclude that the rate equations provide a very useful tool to analyze the non-equilibrium kinetics during the first few picoseconds. The temporal evolution of the population of different states can be successfully described by the net fluxes in terms of a set of statistically averaged parameters -- the transition probabilities $T_{\al\be}(t)$. These transition rates can be accurately extracted from the MD data by analyzing the 
temporal behavior of the particle trajectories. 
Typically, we observe that the saturation of the transition rates at their 
equilibrium values can be reached within $t^E \sim 3t_0 - 6t_0$, i.e., $20-40$\,ps,  
for incident energies below $100$\,meV. The equilibration takes longer for larger incident angles/energies and lower lattice temperatures. 

\subsection{Temperature dependence of the transition rates}\label{ss:temperature}
\begin{figure}
  \begin{center} 
  \hspace{-0.5cm}\includegraphics[width=0.5\textwidth]{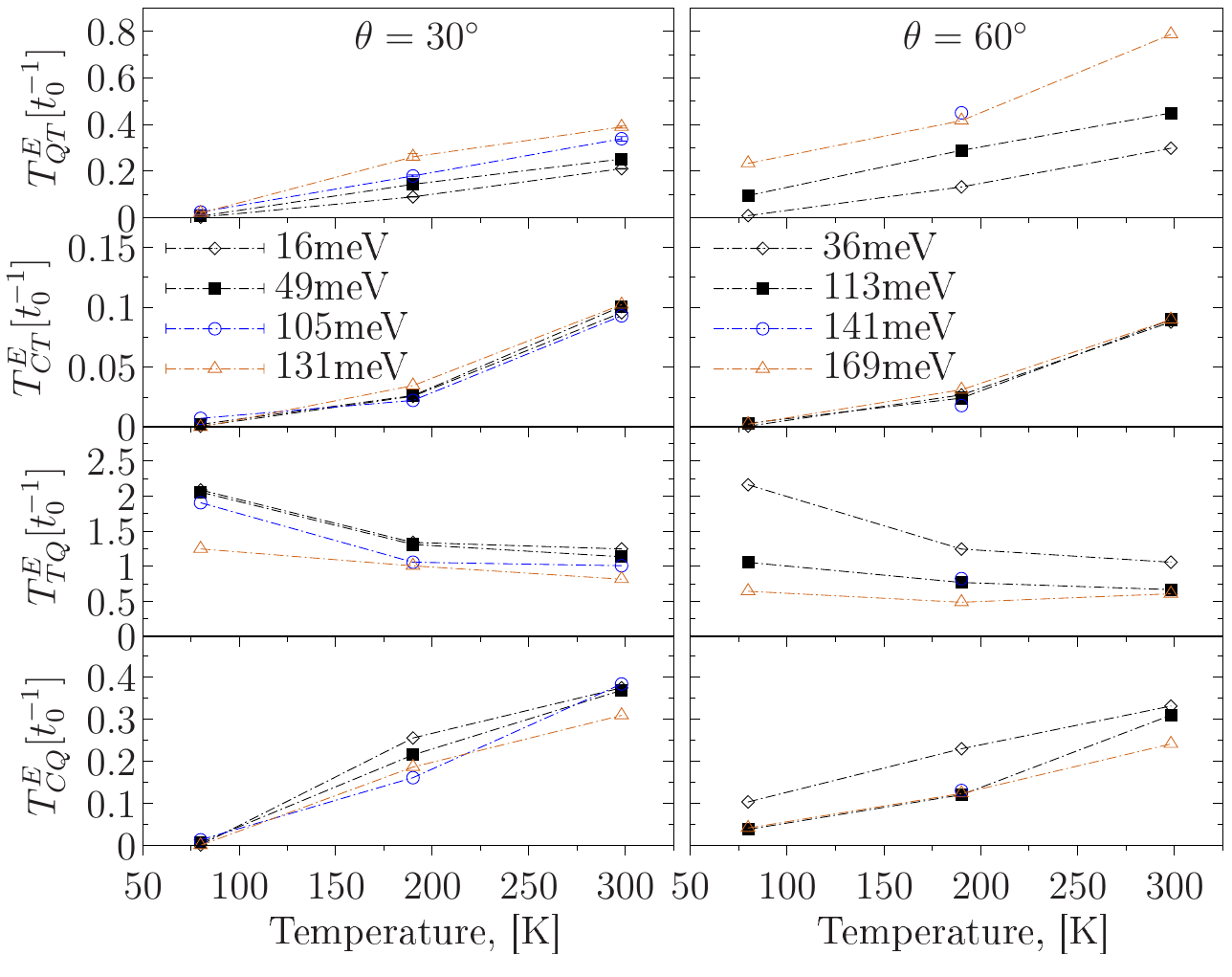}
  \end{center}
  \vspace{-.80cm}
  \caption{Equilibrium transition rates $T_{\al\be}^E$ between different states [trapped ($T$), quasi-trapped ($Q$) and continuum ($C$)] for different incident energies $E_i$
  at the three lattice temperatures $T_s=80$, $190$ and $300$\,K and two incident angles $\theta=30^\circ$ (left panel) and $60^\circ$ (right panel).}
  \label{fig:ratesag3060comp}
  \end{figure}

%
Results for the transition rates at quasi-equilibrium for various incidence conditions, 
i.e., different incident energies $E_i$, incident angles $\theta$ and surface temperatures $T_s$, 
are summarized in figure~\ref{fig:ratesag3060comp}.

First, we discuss the de-excitation transition from Q to T states (Q$\rightarrow$T). The corresponding rate $T^E_{TQ}$
exhibits only a weak dependence on temperature and incident angle/energy. 
It 
stays practically unchanged 
for incident energies $E_i \leq 130$\,meV and $\theta=30^{\circ}$
when  the lattice temperature is varied in the range $80\,K\leq T_s\leq 300\,K$. 
Hence, the binary atom-atom collisions play  a major role here, 
while surface  temperature effects are secondary. 

The value $T^E_{TQ}$ increases when the incident angle is closer to the surface normal. 
It is about 
$20-30\%$ larger for $\theta=30^\circ$ than for $\theta=60^\circ$. 
However, this 
 comparison  is performed at different incident energies 
 to guarantee that the initial sticking probability is the same for both the incident angles.  
 If the rates $T^E_{TQ}$ are compared at similar incident energies, 
  the observed difference is reduced. 
 This becomes obvious e.g.\ when comparing the 
 cases $E_i=49$\,meV at $30^\circ$ versus  $36$\,meV at $60^\circ$ 
 or $131$\,meV at $30^\circ$ versus $141$\,meV at $60^\circ$ 
 in figure~\ref{fig:ratesag3060comp}. 
 A more detailed analysis of the ($\theta,E_i$)-dependence would require a larger set of data.

Now, we consider the transition rates to a higher energy state: T $\rightarrow$ Q, T $\rightarrow$ C and Q $\rightarrow$ C. They reveal a strong, partially 
linear dependence on the lattice temperature. 
In addition, the rate $T^E_{QT}$ shows a clear dependence on the initial energy $E_i$, whereas the rate $T_{CT}$ 
is almost independent of $E_i$. 
Such behavior originates from a different portion of energy transferred from the lattice and 
required for each type of excitation. Much less energy is required for the T $\rightarrow$ Q transition, when only the parallel component of kinetic energy needs to be changed to make the total  energy positive. 
The most probable contribution is expected from the trapped states in the high-energy tail of the distribution function. Here, some correlations with $E^{\parallel}_{k}$ in the incident beam should be present. 
In contrast, during the  T $\rightarrow$ C excitation 
the normal component $E^{\perp}_k$ needs to be changed significantly by an amount comparable with the depth of the physisorption well $\abs{E_0}$ being about 80\,meV for Ar on Pt(111)  
to bring a particle to the continuum. 
Here, the correlations with the incident energies below $\abs{E_0}$ should be small. Note also that the transition probability $T^E_{CT}$ is a factor of $3-4$ smaller than that of $T^E_{QT}$ and $T^E_{CQ}$ 
on the average. Therefore, the most probable excitation to the continuum is a two stage process: T $\rightarrow$ Q followed by Q $\rightarrow$ C.

The transition rates presented in figure~\ref{fig:ratesag3060comp} have a direct practical application.
In combination with the rate equations, they can be used to extrapolate the 
temporal evolution of the state populations to longer time scales being not accessible by usual MD simulations. In particular, this is important for the analysis of thermal desorption at low lattice temperatures, 
such as $T_s=80$\,K, 
when a significant depletion of the trapped states can be observed only on the time scales exceeding those
that are used in the present simulations, i.e., for $t\geq 6 t_0$ (40\,ps). Some applications of this idea will be discussed more in detail below.

\subsection{Analytical solution of the rate equations}\label{ss:analytics}   

In the present section we present the analytical solution of the rate equations  introduced in section~\ref{ss:reconstruction} and demonstrate its efficiency to predict the temporal evolution for longer times. In the end, we derive the estimator of the average residence time of the adsorbate atoms trapped on the surface prior to their thermal desorption.

In case the transition rates are time-independent, the rate equation model reduces to a system of homogeneous linear differential equations of first order and can be solved analytically. 
Therefore, we rewrite Eqs.~(\ref{eq1})-(\ref{REQeq2}) in matrix notation according to 
\begin{eqnarray}
\frac{\db \vec N(t)}{\db t}= \vec R \cdot \vec N(t).  
\label{mrate}
\end{eqnarray}
Here $\vec N(t)=\{N_1(t),N_2(t)\}$ is a column vector with the elements $N_1(t)=N_{T}(t^E+t)$, $N_2(t)=N_{Q}(t^E+t)$, and $\vec R$ is a $2\times 2$ matrix of the transition rates with the elements
\begin{eqnarray}
&R_{11}=-(T_{CT}+T_{QT}), \quad R_{12}=T_{TQ},\nonumber\\
&R_{22}=-(T_{CQ}+T_{TQ}),\quad R_{21}=T_{QT}.
\label{def}
\end{eqnarray}
Notice that 
the fraction of continuum states is not considered here as it follows 
directly from particle number conservation. 

Once the matrix 
$\vec R$ 
is diagonalized, the eigenvalues $\{\lambda_i\}$ and the eigenvectors $\{\vec n_i\}$ 
with $i=1$, $2$ define the complete solution which can be written in the form
\begin{eqnarray}
 \vec N(t)=\sum_{i=1}^2 C_i e^{\lambda_i t} \vec n_i.
 \label{eq:NtMatrixsolution}
\end{eqnarray}
The solution of the eigenvalue problem is given by
\begin{eqnarray}
\lambda_{1(2)} &= -\frac{1}{2} \bigg[\abs{R_{11}}+\abs{R_{22}} \bigg.   \nonumber\\
\quad \bigg.
&\quad \mp \sqrt{(\abs{R_{22}}-\abs{R_{11}})^2+4 R_{12}R_{21}}) \bigg],\\
\vec n_1 &=(\lambda_1-R_{22},R_{21}), \quad \vec n_2=(\lambda_2-R_{22},R_{21}).
\nonumber
\end{eqnarray}
Here, we used explicitly the fact that the diagonal elements are negative, $R_{11}=-\abs{R_{11}}$ and $R_{22}=-\abs{R_{22}}$, as it follows directly from the definition~(\ref{def}).

The expansion coefficients $\{C_i\}$ 
in (\ref{eq:NtMatrixsolution}) can be found by inverting the initial conditions
\begin{eqnarray}
 &\vec N(0)=\sum\limits_{i=1}^2 C_i \vec n_i,\nonumber\\
 &N_1(0)=N_T(t^E), \; N_2(0)=N_Q(t^E) \, .
\end{eqnarray}
They depend on the 
population 
of the trapped and quasi-trapped states at the quasi-equilibration time $t^E$ 
and read 
\begin{eqnarray}
C_{1(2)}=\pm \frac{1}{\lambda_1-\lambda_2}\left(N_1(0)-(\lambda_{2(1)}-R_{22})\frac{N_2(0)}{R_{21}} \right). 
\nonumber
\end{eqnarray}
We summarize the result by writing the complete temporal evolution in the form
\begin{eqnarray}
&N_1(t)=C_1 (\lambda_1-R_{22})e^{\lambda_1 t} +C_2 (\lambda_2-R_{22})e^{\lambda_2 t},\nonumber\\
&N_2(t)=C_1 R_{21}e^{\lambda_1 t}  + C_2 R_{21}e^{\lambda_2 t}.
\label{gensol}
\end{eqnarray}
These results can be further simplified by taking into account that the transition rate $T_{TQ}$ has the largest value (cf.\ figure~\ref{fig:ratesag3060comp}). The relation $T_{TQ} \gg T_{CQ},T_{CT},T_{QT}$, in its turn, leads to $\abs{R_{22}} \gg \abs{R_{11}},R_{12},R_{21}$ which allow us to further simplify the solution by an expansion in the small parameter $g=R_{ij}/\abs{R_{22}}$. 
When considering only the leading terms, the eigenvalues and eigenvectors become
\numparts
\begin{eqnarray}
 &\lambda_{1(2)} =-\left(\abs{R_{11(22)}}\mp \Delta\right)+O(R_{22}\cdot g^4),\\
 & \Delta \lambda= \abs{R_{22}}-\abs{R_{11}}+2\Delta,\quad \Delta=\frac{R_{12}R_{21}}{\abs{R_{22}}},\\
 & \vec n_1 =\left(\Delta \lambda-\Delta,R_{21}\right),\quad  \vec n_2 =\left(-\Delta,R_{21}\right),\\
 &C_1=\frac{1}{\Delta \lambda} \left[ N_1(0)+\Delta \frac{N_2(0)}{R_{21}}\right],\\
 &C_2=-\frac{1}{\Delta \lambda} \left[ N_1(0)-(\Delta\lambda-\Delta)\frac{N_2(0)}{R_{21}}\right].
\end{eqnarray}
\endnumparts

\begin{figure}
  \begin{center} 
  \hspace{-0.4cm}\includegraphics[width=0.52\textwidth]{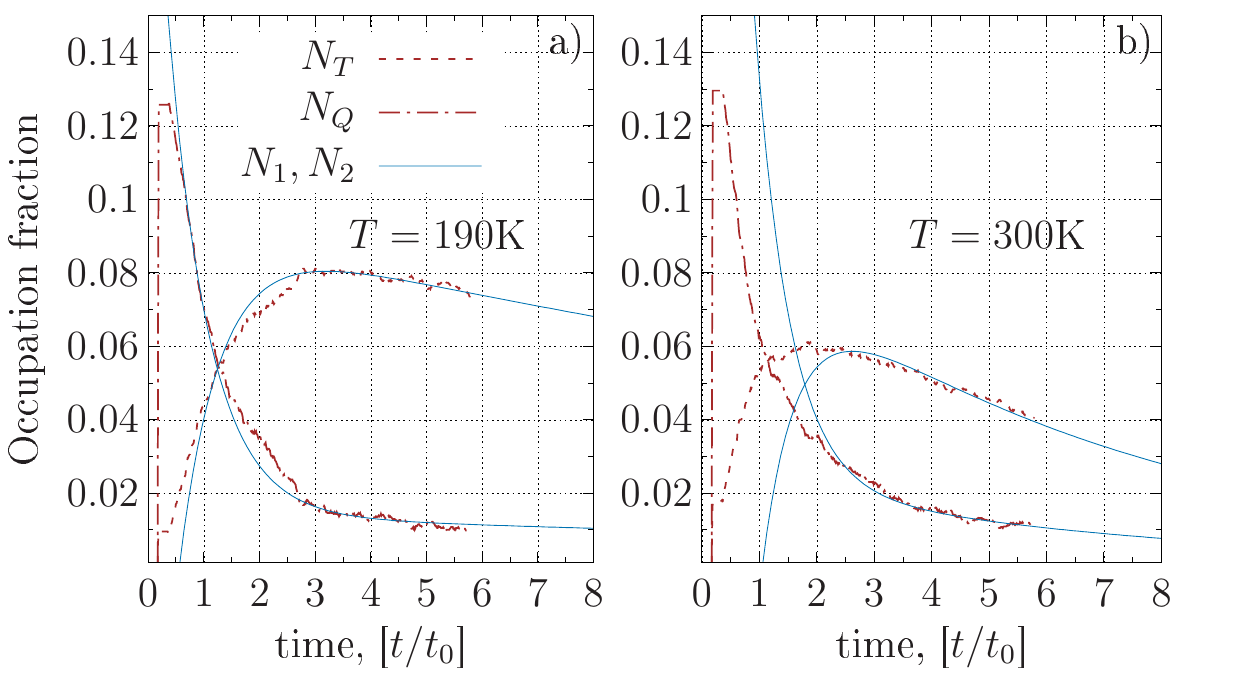}
  \end{center}
  \vspace{-.90cm}
  \caption{{\bf a)} Comparison of the analytical solutions (solid lines), $N_i(t)$, [Eqs.~\ref{gensol} or~(\ref{as2})] with the MD results for $N_T(t)$ and  $N_Q(t)$ for the lattice temperature $T_s=190$\,K. {\bf b)} The same for  $T_s=300$\,K. Incidence conditions:  $\theta=30^\circ$ and $E_i=105$\,meV. In both cases the initial populations, $N_T(t^E),N_Q(t^E)$, are specified at $t^E=3 t_0 = 19.6$\,ps. The analytical solution 
  remains valid for $t\geq t^E$.}
  \label{fig:fig10}
  \end{figure} 
and the dynamics of $N_{1,2}(t)$ are given by 
\begin{eqnarray}
N_1(t)=&N_1(0)\left[(1-\gamma)e^{\lambda_1 t}+\gamma e^{\lambda_2 t}\right]\nonumber\\
&+(1-\gamma)\frac{R_{12}}{\abs{R_{22}}} N_2(0)\left[ e^{\lambda_1 t}-e^{\lambda_2 t}  \right]\nonumber,\\
N_2(t)=&N_2(0)\left[\gamma e^{\lambda_1 t}+(1-\gamma) e^{\lambda_2 t}\right]\nonumber\\
&+ \frac{R_{21}}{\Delta \lambda} N_1(0) \left[ e^{\lambda_1 t}-e^{\lambda_2 t}  \right].
\label{n2a}
\end{eqnarray}
The leading terms are easily identified by the smallness of the parameter $\gamma=(\Delta/\Delta \lambda)\ll 1$.

Using this representation (\ref{n2a}), we  analyze  both the short-time and the long-time behavior. 
The expansion around the initial time $t=0$ yields 
\begin{eqnarray}
&N_1(t)= N_1(0)+R_{11}N_1(0)\cdot t+ R_{12}N_2(0)\cdot t,\nonumber\\
&N_2(t)= N_2(0)+R_{22}N_2(0)\cdot t +R_{21}N_1(0)\cdot t \, ,
\end{eqnarray}
showing that the populations increase linearly both with time and the initial rate values. 

To analyze the long-time behavior, we order the eigenvalues, $\abs{\lambda_1}<\abs{\lambda_2}$ 
which are assumed to be real and negative, $\lambda_i=-\abs{\lambda_i}$.
Due to the exponential decay, the transient processes characterized by 
$\lambda_2$ 
vanish for long times, and  the asymptotic behaviour is governed by the smallest eigenvalue, $\lambda_1$. 
Thus,  we obtain from Eqs.~(\ref{n2a}) the result in the limit $t \gg (\abs{\lambda_2}-\abs{\lambda_1})^{-1}$,
\begin{eqnarray}
&N_1(t)=(1-\gamma)e^{\lambda_1 t}\left[N_1(0)+\frac{R_{12}}{\abs{R_{22}}} N_2(0)\right]\nonumber,\\
&N_2(t)=e^{\lambda_1 t}\left[N_2(0) +\frac{R_{21}}{\Delta \lambda} N_1(0)\right]\,. 
\label{as2}
\end{eqnarray} 
Consequently, both states decay with the same exponential factor and satisfy a more general detailed balance equation that can be derived directly from Eq.~(\ref{gensol}), without any approximation,
\begin{eqnarray}
 R_{21} \cdot N_1(t)\approx (\lambda_1-R_{22}) \cdot N_2(t).
 \label{db1}
\end{eqnarray}

Now we take into account that $\Delta, \abs{R_{11}} \ll \abs{R_{22}}$ and $\lambda_2\approx \Delta \lambda -\Delta\approx \abs{R_{22}}$. Using the definition~(\ref{def}) and Eq.~(\ref{db1}) we obtain the detailed balance equation
\begin{eqnarray}
 T_{QT} \cdot N_{T}(t)\approx (T_{TQ}+T_{CQ}) \cdot N_{Q}(t)\,,
 \label{db2}
\end{eqnarray}
showing that the decay of the quasi-trapped states, $Q\rightarrow C$ and $Q\rightarrow T$, is balanced by the thermal excitation, $T\rightarrow Q$, of the trapped states. 
This fact also confirms that the ratio $N_{Q}(t)/N_{T}(t)$ saturates 
 and is uniquely determined by the transition rates in the quasi-equilibrium phase. %

To verify the validity of the analytical solution 
(\ref{n2a}) for $N_{1(2)}(t)$, we compare them with the MD results,  $N_{T(Q)}(t)$,  in Fig.~\ref{fig:fig10}. 
Obviously, the agreement is very good, for $t\geq t^E$. This confirms the main advantage of the rate equation model--its simplicity and ability to provide explicit results at arbitrary times beyond $t^E$.

As a second important application of the analytical solution 
(\ref{n2a}), we estimate the residence time, $t^R$, which characterizes how long the adsorbate atoms stay trapped near the surface prior to thermal desorption. The derived asymptotic behavior yields
\begin{equation}
 t^R_{\lambda_1} = \abs{\lambda_1}^{-1}.
 \label{trr}
\end{equation}
A more quantitative definition of $t^R$ follows from the condition that  the adsorbate concentration is reduced by some factor $\nu$ during the time interval $[t^E, t^R]$. Using the asymptotic limit of both concentrations~(\ref{as2}), the residence time can be defined solely in terms of the initial concentrations and the stationary transition rates
\numparts
\begin{eqnarray}
 t^R=&\frac{1}{\lambda_1} \ln 
\left( \frac{\gamma \left[N_1(0)+N_2(0)\right]}{N} \right)
 ,\label{eq:TRapprox}\\
 N=&(1-\gamma)\left[N_1(0)+\frac{R_{12}}{\abs{R_{22}}} N_2(0)\right] 
\nonumber
\\
 &+ \left[N_2(0) +\frac{R_{21}}{\Delta \lambda} N_1(0)\right]. 
 \label{eq:Napprox2}
\end{eqnarray}
\endnumparts
Note that the ratio of concentrations in the logarithmic term is typically of the order of one 
so that the estimate~(\ref{trr})  results again from (\ref{eq:TRapprox}) with $\gamma=e^{-1}$. 

Finally, we summarize our results for $t^R$, and its dependence on the incidence conditions and the lattice temperature in table~\ref{tresidence}. As expected,  $t^R$
 depends only weakly on the incidence conditions  for the same lattice temperature, where  
 the variations are within the statistical errors.
 This confirms that any memory of the incidence conditions is lost within 
 $t^E = 3 t_0 \approx 20$\,ps, 
 and the analytical solutions~(\ref{as2}), constructed for $t\geq t^E$, accurately describe both the decay of the adsorbate and its characteristic residence time. 
Our simulations predicted that  
the residence times are $t^R=(170-180)$\,ps for $T_s=190$\,K and $(50-55)$\,ps for $T_s=300$\,K.

For $T_s=80$\,K, we need to extend our simulations beyond $t=6 t_0 \approx 40$\,ps to provide a more accurate estimation of the transition rates $T_{ij}^E$ and 
 $t^R$. A noticeable decay of the trapped fraction due to  thermal desorption just starts at 40\,ps and, therefore, the simulations need to 
 be extended to at least  100\,ps to ensure that the constructed analytical solution fits well the MD simulation data 
 similar to the cases presented in figure~\ref{fig:fig10}.  
Still some estimate can be given based on the parametrization of the experimentally determined desorption times by a Frenkel-Arrhenius formula
\begin{equation}
t^R = t_p \,e^{U/(k_B T)}.
\label{eq:tr_model}
\end{equation}
Here the prefactor $t_p$ typically varies for physisorbed gases from $10^{5}$\,ps for helium desorbing 
from constantan~\cite{td1} to $10^{-2}$\,ps for xenon desorbing from tungsten\cite{td2}. By expressing the adatom desorption frequency as $\nu=1/t^R$, the prefactor $t_p$ can be interpreted as an average time between the successive bounces on the surface, and the Boltzmann factor $e^{-U/(k_B T)}$ as the static desorption probability, with $U$ being comparable with the depth of the surface potential $E_0$. We found that such interpretation applies very reasonable to our system. We estimated $t_p=1.52$\,ps and $U=61.7$\,meV, by applying the  fit~(\ref{eq:tr_model}) to our data for $t^R$ at $T_s=190$\,K and $300$\,K (see table~\ref{tresidence}). The temperature $T$ was chosen to be the effective adsorbate temperature $T^\star$~\cite{paper2}, where we 
 have used  $T^\star=150$\,K for $T_s=190$\,K and 200\,K for $T_s=300$\,K. The obtained fit parameters well agree with the  depth of physisorption potential, $\abs{E_0} \sim 78$\,meV (see table~\ref{tab1}), and the average time between the bounces, which varies in the range from 0.93\,ps to 1.10\,ps for the lattice temperature $80\,\text{K}\leq T_s\leq 300\text{K}$. Here, we have used the time dependence of average bounce number $\avr{n_b}(t)$, presented in figures~\ref{fig:ntypeT26T10ag30e20} and ~\ref{fig:ntypeT06ag60}.

Finally, the extracted fit parameters, $t_p$ and $U$, allow to estimate the residence time, $t^R \sim 11000$\,ps, at the lattice temperature $T_s=80$\,K [using $T^\star=80$\,K].  This value agrees quite well with the lower bound for $t^R$ presented in table~\ref{tresidence}. 

\begin{table}
\caption{Residence time of Ar on Pt(111) estimated from Eq.~(\ref{eq:TRapprox}) [with $\gamma=e^{-1}$] for different incidence conditions ($\theta$ and $E_i$) and lattice temperature $T_s$. The last column shows the deviation from a more simple estimate~(\ref{trr}).}
  \label{tresidence}
 \begin{tabular}{c c c c c}
 \hline
 \hline
$T_s$ & $\theta$ & $E_i$[meV] & $t^R$[ps] &$t^R_{\lambda_1}/t^R$ \\
 \hline
80\,K      & $30^\circ$& 15.9& >7900 & 0.99\\ 
           & $60^\circ$& 36.3& >8000 & 0.98\\
\hline
190\,K      & $0^\circ$& 12.8& 184(10) & 0.95\\
            & $30^\circ$& 15.9& 179(10) & 0.95\\
            &           & 49.4 &165(10) & 0.92 \\
            &           & 105.3 &182(10) & 0.88 \\
      & $60^\circ$& 36.3& 163(10) & 0.91\\
            &           & 112.7&169(10) & 0.80 \\
\hline
300\,K      & $0^\circ$& 12.8& 57(5) & 0.90\\
            & $30^\circ$& 15.9& 56(5) & 0.90\\
	    &           & 49.4 &52(5) & 0.87 \\
	    &           & 105.3 &50(5) & 0.85 \\
	    &           & 131.1 &49(5) & 0.85 \\
      & $45^\circ$& 21.6& 55(5) & 0.90\\
      &   & 100.7& 46(5) & 0.82\\
      & $60^\circ$& 36.3& 56(5) & 0.86\\
	    &       & 112.7& 50(5) & 0.78 \\
\hline
\hline
 \end{tabular}
 \end{table}

\section{Conclusion}\label{s:conclusion1}

The studied kinetics of adsorption and desorption of atomic projectiles physisorbed on solid metallic surfaces is the most elementary process serving as a starting point for a detailed understanding of more complex processes, i.e. for the chemisorbed species which can undergo substantial structural and electronic modifications. In the present study, we restricted ourselves to  physisorption at low coverage so that the interaction between gas particles in the adsorbate can be neglected.

Our main motivation was to explore the capabilities of accurate MD simulations for the sticking of argon atoms on a metal surface. However, the main obstacle is the enormous difference in the time scales of the atomic motion being about $10^{-13}$\,s which has to resolved in the simulations and of the desorption processes ranging  from $10^{-6}$ to few seconds 
in experiments. These scales cannot be reached with MD simulations, even on supercomputing hardware, without further approximations. 
Therefore, we developed a new approach that couples MD simulations to an analytical rate equations model which has  allowed us to extend the calculations to 
several hundreds of picoseconds, and further extensions are possible as well. The rate equations are not trivial, as one first has to realize that atoms near the surface after the first scattering event have to be classified into three possible categories: continuum (desorbed), quasi-trapped state (moving in the surface plain) and trapped ones. 

Most importantly, we have demonstrated that this combination of MD and rate equations can be performed successfully \textit{without loss of accuracy}, for times exceeding the equilibration time $t^E$. 
The key is that the time $t^E$ and all relevant input parameters to the rate equations--the transition rates between the particle categories--are directly extracted from the MD data for which a reliable procedure has been developed.
The rates are found to have a strong time dependence, during the initial period,
until they saturate for times $t$ around the equilibration time $t^E$, after which they remain constant. 
i.e. $T_{\al\be}(t)|_{t>t^E} \approx T_{\al\be}^E$. 
These stationary values are 
determined by the shape of the quasi-equilibrium energy distribution function of the atoms in contact with the surface which
 is discussed in detail in paper~II~\cite{paper2}. 

Let us now critically discuss limitations and possible improvements.
First, our statistical approach, of course,  does not contain a microscopic treatment of the individual quantum scattering events. The dynamics were treated semi-classically by adopting binary interaction potentials that reconstruct the potential energy surface precalculated by state-of-the-art DFT calculations~\cite{Leonard}.
Further improvements are possible by using 
more accurate force-fields in the MD simulations. 

Second, the derivation of the rate equations in Sec.~\ref{s:derivation} was based on several assumptions: our starting equation (\ref{Tfi})  for the transition probability  was based on a standard Markovian approximation for the interaction with the dissipative subsystem. Specifically, we ignored possible correlations between different adsorbate states, i.e., the memory effects for interstate transitions. Next, we have assumed that the phonon-induced transitions dominate, while thermal excitations of the solid are rapidly dissipated due to fast vibrations and the coupling of the substrate atoms. At the same time, the excellent agreement with the MD simulations provides strong support for these assumptions. 
On the other hand, we underline, that our rate equations are valid only at low coverage with adsorbate atoms. 
At higher coverage, surface states will be blocked by adsorbed atoms. These effects can be straightforwardly included into the rate equations which then become nonlinear in the concentration. Such extensions will be presented in a future study.

Third, the rate equations description becomes very efficient once the total system of gas plus surface has reached thermal equilibrium. The main requirement is that the relaxation time for an atom on the surface to reach local equilibrium is much less than the typical residence time of an atom on the surface. In this regime, the gas atoms lose memory of their initial state and become randomized with respect to energy and momentum. 
Interestingly, this situation is particularly well fulfilled at low lattice temperatures, $T_s \leq 80$K, when the thermal desorption is extremely slow presenting a challenge to MD simulations. The particle trajectories must then be integrated during a very long residence time. Moreover, to obtain a good statistics over the desorption rates, the angular and velocity distributions, thousands of trajectories must be sampled. 
To make the problem tractable, 
 techniques for treating ``rare events'' have been proposed~\cite{sto1,depristo}, for an overview see Ref.\cite{bonitz_psst_18}. In particular, 
studies of the thermal desorption of Ar and Xe from Pt(111) have been conducted by means of the stochastic classical trajectory approach~\cite{meta1,meta2} 
when the residence time exceeds 1\,s.


Finally, the present simulations did only consider the scattering of single atoms, one at a time. In the case of plasmas in contact with a surface this is justified at sufficiently low pressures and particle fluxes to the surface. As a consequence,  at long times the fraction of trapped and quasi-trapped atoms is slowly decreasing, cf. Fig.~\ref{fig:fig10}. For the computation of the sticking probability \cite{paper2} and of the residence time these time scales are not essential. However, for other applications such as the growth dynamics of an adsorbate layer the long-time behavior is of direct interest. In that case it is expected that the decay of the adsorbed fractions is compensated by the continuous influx of atoms from the plasma leading to a quasi-stationary state. The present rate equations model can be straightforwardly extended to include this flux as a source term. This will enable one to study these processes systematically in dependence on the plasma conditions.


%


\end{document}